%% file: main.tex
\documentclass[lettersize,journal]{IEEEtran}
\input{macros}

\begin{document}

\title{\name: Evaluating Large Language Models for Implicit and Explicit Code Execution Reasoning}

\author{Changshu Liu, Yang Chen, Reyhaneh Jabbarvand 
\thanks{University of Illinois Urbana-Champaign}
\thanks{Emails:\{cl144, yangc9, reyhaneh\}@illinois.edu}
}

\markboth{IEEE TRANSACTIONS ON SOFTWARE ENGINEERING,~Vol.~x, No.~x, January~2026}
{Shell \MakeLowercase{\textit{et al.}}: A Sample Article Using IEEEtran.cls for IEEE Journals}

\maketitle

\begin{abstract}
Large Language Models (LLMs) have been widely used to automate programming tasks. Their capabilities have been evaluated by assessing the quality of generated code through tests or proofs. The extent to which they can reason about code is a critical question revealing important insights about their true capabilities. This paper introduces \name, a framework designed to gauge the code reasoning abilities of LLMs through the following explicit and implicit code reasoning tasks: \textit{\underline{I}ndependent \underline{E}xecution \underline{R}easoning (\IER)}, \textit{\underline{S}pecification \underline{R}easoning (\SR)} and \textit{\underline{D}ynamic \underline{S}emantics \underline{R}easoning (\CR)}. The first evaluates the abilities of LLMs to simulate the execution of given inputs to a code and predict the output (\IER). The second assesses the abilities of LLMs to incorporate the simulation of test data in the specification into code generation (\SR). Finally, \name evaluates LLMs' abilities to understand overall code semantics only given a specific input/output (\CR). 

Our extensive evaluation of \major{$13$} LLMs across four widely used benchmarks using \name shows that LLMs, depending on their size and training strategy, can reason about some dynamic aspects of code. However, their performance drops for code with higher complexity, nested code constructs, non-primitive types, and intra-class dependencies. We show that these reasoning tasks evaluate LLMs differently, and a comprehensive evaluation of code reasoning requires them all. Finally, we show that the performance of LLMs in bug repair is not correlated with any of the code reasoning tasks, and except for advanced frontier models, other LLMs do not incorporate code reasoning when performing bug repair. Given that program repair requires execution reasoning (to determine where the behavior of buggy code differs from specified behavior to localize the bug) as well as specification and dynamic semantics reasoning (to re-write the code such that the patch keeps correct semantics but fixes semantic mismatch with the specification), this observation raises the question of to what extent we can trust these models for programming tasks that require code understanding and analysis.  
\end{abstract}

\vspace{-10pt}
\begin{IEEEkeywords}
Code Reasoning, Large Language Models, Program Repair
\end{IEEEkeywords}

\input{sections/1-Introduction}

\input{sections/2-Approach}

\input{sections/3-Setups}

\input{sections/4-Evaluation}

\input{sections/5-Related-Work}
\input{sections/6-Conclusion}

\bibliographystyle{IEEEtran}
\bibliography{acmart}


\newpage

\end{document}

%% file: macros.tex
\usepackage{graphicx}
\usepackage{amsmath} 
\usepackage{stmaryrd}
\usepackage{xspace}
\usepackage{cleveref}
\usepackage{multirow}
\usepackage{nth}
\usepackage{wrapfig}
\usepackage[font=small,labelfont=bf]{caption}
\usepackage{subcaption}
\usepackage{enumitem}
\usepackage{bm}
\usepackage{soul,color}
\usepackage[table,xcdraw]{xcolor}
\usepackage{listings}
\usepackage{verbatim}
\usepackage{ragged2e}
\usepackage{float}
\usepackage{xurl}
\usepackage{colortbl} 
\usepackage{amssymb}
\IEEEaftertitletext{\vspace{-1cm}}

\usepackage{tabularx,array}
\usepackage[dvipsnames]{xcolor}

\newcommand{\revision}[1]{\textcolor{black}{#1}}
\newcommand{\major}[1]{\textcolor{black}{#1}}

\newcommand{\omini}{o4-mini\xspace}
\newcommand{\dsr}{DeepSeek-R1\xspace}

\newcommand{\gptf}{GPT-4\xspace}

\newcommand{\gemini}{Gemini-1.5-Pro\xspace}
\newcommand{\claude}{Claude-Sonnet-4.6\xspace}
\newcommand{\deepseek}{DeepSeekCoder\xspace}
\newcommand{\codellama}{CodeLlama\xspace}

\newcommand{\starcoder}{StarCoder\xspace}

\newcommand{\sem}{SemCoder-S\xspace}
\newcommand{\heval}{HumanEval\xspace}

\newcommand{\avatar}{Avatar\xspace}
\newcommand{\crux}{CRUXEval\xspace}
\newcommand{\ceval}{ClassEval\xspace}
\newcommand{\name}{CodeMind\xspace}
\newcommand{\starT}{StarCoder\,2\xspace}
\newcommand{\reval}{\textsc{REval}\xspace}
\newcommand{\refactor}{Dynamic Semantics Reasoning\xspace}
\newcommand{\CR}{DSR\xspace}
\newcommand{\IER}{IER\xspace}
\newcommand{\SR}{SR\xspace}

\newcommand{\exerscope}{ExeRScope\xspace}
\newcommand{\greenuparrow}{\textcolor{ForestGreen}{\uparrow}}

\definecolor{ForestGreen}{RGB}{34,139,34}

\newcommand{\coco}{\textsc{CocoNut}\xspace}

\definecolor{codegreen}{rgb}{0,0.6,0}
\definecolor{codegray}{rgb}{0.5,0.5,0.5}
\definecolor{codepurple}{rgb}{0.58,0,0.82}
\definecolor{backcolour}{rgb}{0.95,0.95,0.92}
\definecolor{lightgray}{gray}{0.9}

\lstdefinestyle{stylewithcommentPy}{
language=Python,
basicstyle=\linespread{0.9}\scriptsize,
backgroundcolor=\color{backcolour}, commentstyle=\color{codegreen},
keywordstyle=\color{magenta},
numberstyle=\tiny\color{codegray},
basicstyle=\scriptsize\ttfamily,
escapechar=|,
stringstyle=\color{codepurple}, 
xleftmargin=0cm, 
frame=tlbr, 
framesep=0.1cm, 
framerule=0pt, 
numbers=none, 
breaklines=true,
moredelim=**[is][\color{red}]{@}{@},
moredelim=**[is][\color{black}]{`}{`},
moredelim=[is][\color{red}\bfseries\underbar]{~}{~},
moredelim={[is][\color{black}\bfseries]{@@}{@@}}
}

%% file: sections/1-Introduction.tex
\vspace{-10pt}
\section{Introduction}
\label{sec:introduction}

Large Language Models (LLMs) have shown emerging abilities in automating different programming tasks. However, several studies suggest they struggle to generalize this ability to real-world programs~\cite{du2023classeval,jimenez2023swe} or to tasks that require understanding code logic rather than natural language~\cite{pan2023understanding,min2023beyond}. This is mainly because LLMs are trained to associate code generation with natural language specifications, i.e., combine code constructs similar to thousands to millions of examples they have seen while aligning to the requirements specified in the natural language. As a result, they inherently have limited abilities to perform broader program analysis tasks or perform reliably when natural language hints do not exist.

A large body of work has assessed LLMs for reasoning tasks of different modalities \cite{deshpande2021rect,wu2023reasoning,miceli2023larger,bubeck2023sparks,wang2023mathcoder,imani2023mathprompter,luo2023wizardmath,huang2023lvlms,valmeekam2022large,min2023beyond}, including natural language, visual data, math, and logic. Recently, code reasoning, \revision{the ability of the models to understand code semantics and incorporate that in their decision-making}, has become a popular evaluation strategy for assessing LLMs.
\crux~\cite{gu2024cruxeval} is a benchmark of synthetically generated simple Python programs and corresponding input/output pairs, focusing on evaluating the abilities of LLMs in input and output predictions. \reval~\cite{chen2024reasoning} is a framework to assess the abilities of LLMs in predicting dynamic execution properties such as output prediction, branch prediction, and intermediate variable value prediction. 
\revision{Beyond explicitly focusing on code execution, there also exist tasks that require LLMs to \emph{implicitly} apply their understanding of code execution. 
Without such reasoning, correct code produced by LLMs may simply reflect overfitting to benchmarks, superficial pattern matching, or data contamination. For instance, in code synthesis, models are always provided with test cases that serve as specifications of the expected input-output behavior. Without an accurate understanding of how the test cases would be executed, it is unlikely that the model can synthesize code that fully satisfies the intended functionality. This emphasizes the importance of execution-aware reasoning, even in tasks where such reasoning is not explicitly prompted.}

\begin{figure}[H]
\vspace{-10pt}
    \centering
    \includegraphics[width=\linewidth]{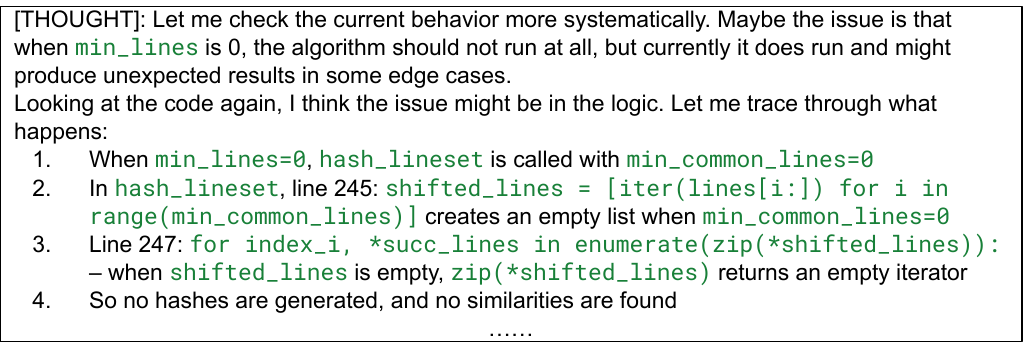}
    \caption{\major{The reasoning process of SWE-agent (Claude-Sonnet-4) at trajectory step 42 for pylint-4970}}
    \label{fig:agent_thought}
    \vspace{-10pt}
\end{figure}

\major{Agentic systems, although equipped with external tools, still may need to reason about code execution to properly localize or edit a patch, prior to any tool usage. In Figure~\ref{fig:agent_thought}, we present an example\footnote{The complete trajectory can be found in \cite{sweagent_traj}.} where SWE-agent~\cite{yang2024swe} internally simulates the program’s execution behavior before applying a code edit, even though such step-by-step execution reasoning \emph{is not explicitly requested} by the system prompt.}
\major{In fact,} none of the prior techniques focus on \emph{implicit} code reasoning, i.e., designing tasks, metrics, and experiments assessing whether LLMs incorporate explicit reasoning about code execution when performing other programming tasks. 

This paper introduces \name framework, which formally defines three explicit and implicit code reasoning tasks and metrics: \textbf{Independent Execution Reasoning (\IER)}, an \emph{explicit} reasoning task that assesses if LLMs can reason how given inputs evolve to output for any arbitrary code. \textbf{Specification Reasoning (\SR)}, an \emph{implicit} reasoning task that evaluates the extent to which LLMs can incorporate the simulation of test data in the specification to generate correct code. \textbf{\refactor(\CR)}, an \emph{implicit} reasoning task that assesses the abilities of LLMs in generalizing the understanding of overall code semantics only given a specific input/output and refactoring it to a shorter, semantically equivalent version when possible. Using \name, we performed a large-scale study to assess state-of-the-art LLMs for code reasoning. We selected \major{\emph{13}} models, including both general-purpose and Code LLMs, and prompted them for IER, SR, and \CR tasks on \emph{1450} programs written in Python. These programs are from \emph{four} programming benchmarks, namely \heval~\cite{humaneval}, \crux~\cite{gu2024cruxeval}, \ceval~\cite{du2024evaluating}, and \avatar~\cite{ahmad2021avatar}. Our framework and experiments answer the following research questions:

\begin{itemize}[leftmargin=*]
    \item \emph{To what extent can LLMs explicitly and implicitly reason about code?}

    \textbf{RQ1: Performance of LLMs in IER.} LLMs can \textit{explain} the code statement by statement and often follow the execution flow. Open-source LLMs that have achieved comparable effectiveness as closed-access models (e.g., \gptf and \gemini) in code synthesis are behind them with a \emph{notable gap} concerning execution reasoning (\S\ref{subsec:eval-IER}).

    \textbf{RQ2: Performance of LLMs in SR.} LLMs, to a limited extent, can reason about test data in the specification and bring that into the code synthesis process. The more ambiguous and non-informative the natural language specification, the more helpful it is to include tests (\S\ref{subsec:eval-SR}).  

    \textbf{RQ3: Performance of LLMs in \CR.} LLMs can understand general code semantics, although to a limited extent, and refactor arbitrary code by removing redundant code constructs (\S \ref{subsec:eval-DSR}). 

    \item \emph{What factors impact the code reasoning abilities of LLMs?}

    \textbf{RQ4: Analysis of Reasoning Failures.} With automated analysis of reasoning failures, accompanied by a detailed, in-depth study of LLM's chain of thought reasoning, we observe that nested code constructs, complex conditional predicates and loop conditions, the non-trivial combination of arithmetic and logic operators, and API invocations can significantly challenge LLMs for explicit and implicit code reasoning (\S \ref{subsec:eval-factors}). 

    \item \emph{Do we need both explicit and implicit code reasoning to evaluate LLMs?}

    \textbf{RQ5: Necessity for Different Code Reasoning Tasks.} LLMs' performance across code reasoning tasks is inconsistent: models may correctly reason about the execution of a test input (\IER) but fail to incorporate the test data when synthesizing the code (\SR). They may also correctly reason about code execution of specific inputs (\IER) and incorporate that into code generation (\SR) but fail to generalize the reasoning about all inputs (\CR). These results entail evaluating LLMs under different reasoning tasks (\S \ref{subsec:eval-different-reasoning}).

    \item \emph{Does a better (explicit or implicit) code reasoning result in better performance in programming tasks, e.g., bug repair?}

    \textbf{RQ6: Association Between Code Reasoning and Program Repair.} Prior to reasoning models, there is no meaningful association between the bug repair abilities of the models and different code reasoning tasks. Even when we instruct LLMs to reason about execution in their chain of thought for performing bug repair tasks, only frontier LLMs, e.g., \claude, \omini, and \dsr, incorporate explicit execution reasoning (\IER) to localize and repair the bug. Others, even when instruction-tuned on execution data, fail to do it by default or follow the instructions. A deep investigation into the cases where LLMs successfully repair bugs but fail to explicitly or implicitly reason about code shows that the success in such cases could be due to natural language shortcuts, lucky hallucinations (potentially due to data leakage), or a high degree of code clones in open-source software, without understanding the nature of the bug. 

    \item \major{\emph{What is the role of code reasoning in programming agents?}}
    
    \major{\textbf{RQ7: Code Reasoning in Programming Agents}. We inspect $2,000$ SWE-agent trajectories and find that $20.95\%$ exhibit execution-aware reasoning before taking actions, using a lightweight keyword-based detection strategy (\S \ref{subsec:agentic}). Subsequent manual analysis further confirms that (implicit) code reasoning is embedded in the problem-solving process of tool-augmented coding agents, helping them avoid repetitive and unproductive actions.}

    \item \major{\emph{How does \name compare with existing code-reasoning evaluation approaches?}}

     \major{\textbf{RQ8: Comparison with Alternative Approaches}. We compare the \IER of \name with \reval, focusing on output prediction for the overlapping subset of \heval and \ceval programs and the common LLMs studied in both techniques (\S\ref{subsec:comparison}). This comparison indicates that \name achieves more correct (and more unique) output predictions, due to its prompt design with explicit instructions and crafted in-context examples.}

\end{itemize}

Our contributions include (1) \name framework defining three explicit and implicit code reasoning tasks; (2) a large-scale evaluation of LLMs for code reasoning using \name; (3) a code reasoning benchmark beyond simple, less diverse, and synthetic programs in \crux, helping generalize the conclusions from observations; (4) a comprehensive, in-depth analysis of results that offers a catalog of root causes negatively impacting the abilities of LLMs for code reasoning; and (5) studying the association between code reasoning and program repair, as a representative programming task that requires both explicit and implicit code reasoning.

%% file: sections/2-Approach.tex
\vspace{-5pt}
\section{\name}
\label{sec:CodeMind}

Program specification (either in natural language, code, or mathematical expressions) defines the logic that the code should implement. Formally speaking, it defines a function $S: S_I \rightarrow S_O$, where $S_I$ is a set of all possible inputs to the program and $S_O$ is a set of corresponding outputs. A code synthesized based on the implementation is also a function $C: C_I \rightarrow C_O$. We define a program to be \textit{correct with respect to specification} if it satisfies all the following conditions:
\begin{center}
$C_I \subseteq S_I$, $C_O \subseteq S_O$, $\forall i \in C_I, C(i) = S(i)$    
\end{center}

If we want models to synthesize a \emph{correct} code (with respect to the provided specification), this entails reasoning about how inputs evolve to outputs through implementation (Independent Execution Reasoning) and implementing the code such that it generates correct output for given inputs (Specification Reasoning). Ultimately, the model should reason about the entire input space and the evolution of individual inputs to their corresponding expected outputs, understanding dynamic code semantics (Dynamic Semantics Reasoning).

\vspace{-5pt}
\subsection{Independent Execution Reasoning}
\label{subsec:ERdefinition}

Considering the formalization above, we define the independent execution reasoning task as follows: 

\vspace{3pt}
\textbf{Definition 1: Independent Execution Reasoning (IER).} 
Given a program $C: C_I \rightarrow C_O$ and set of inputs $\hat{I} = \{i | i \in C_I\}$, LLM $L$ succeeds in IER if $\hat{o} = C(\hat{I})$, where $\hat{o}=L(\hat{I})$ is the predicted output by $L$. Note that in this task, we do not deal with specification, so we can assess LLMs for any arbitrary code with ground-truth pairs of $\langle\hat{I},\hat{o}\rangle$.
IER is an explicit code reasoning task that evaluates LLMs for general inductive code reasoning. Succeeding in this task requires LLMs to know different code constructs, arithmetic and logic operations, and PL-specific properties, e.g., list comprehension and lambda expression in Python. 
\name measures the performance of a model $L$ in IER for a given program $C$ with inputs $\hat{I}$ using the following metric:

\begin{small}
\vspace{-5pt}
\begin{equation}
\label{eq:SIER}
  S_{IER}(L,C,\hat{I})=\begin{cases}
    1, & \text{if $L(\hat{I})=C(\hat{I})$}\\
    0, & \text{otherwise}
  \end{cases}
\end{equation}  
\end{small}

Given that LLMs are mostly evaluated on benchmarks, \name also offers \textit{IER Rate ($R_{IER}$)}, a collective metric that measures how much a given LLM $L$ can reason about multiple programs in a benchmark. \name calculates $R_{IER}$ for a set of $m$ programs in benchmark $B_{|m|}$ as follows:

\begin{small}
\vspace{-5pt}
\begin{equation}
\label{RIER}
   R_{IER}(L,B_{|m|})=\dfrac{\sum\limits_{i=1}^{m} \llbracket S_{IER}(L,C_i \in B,\hat{I_i})=1 \rrbracket}{m} 
\end{equation}
\end{small}

The Iverson bracket $\llbracket$$\rrbracket$ returns $1$ if the condition in square brackets is satisfied and $0$ otherwise.

\vspace{-5pt}
\subsection{Specification Reasoning}
\label{subsec:SRdefinition}

Concerning the generation of correct code, a model should understand specifications to synthesize the correct code. When the specification is in natural language, this can be achieved by instruction-tuning natural language and code generation so that models map the specified concepts in the specification to the sequence of code tokens. The specification can also include test data, e.g., as feedback to LLM for fixing the previously generated incorrect code or enabling test-driven code synthesis. Incorporating more formal information, such as test data, requires a different alignment approach in LLMs. That is, the model should be able to reason about the execution of given inputs and implement the code to yield the same output. We define such an implicit reasoning task as follows:

\vspace{3pt}
\textbf{Definition 2: Specification Reasoning (\SR).}
Given a problem specification $S: S_I \rightarrow S_O$ in natural language, a test $t=\langle i,o \rangle$, where $i \in S_I, o \in S_O, S(i) = o$, program $C_S$ (generated given the specification $S$), and program $C_{S+t}$ (generated given the specification $S$ and test $t$), the LLM succeeds in SR if $C_S(i) \neq o \quad \& \quad C_{S+t}(i) = o$. That is, the LLM that previously was not able to generate a correct code, i.e., the generated program ($C_S$) failed on test suite $T$ , can now generate a correct code ($ C_{S+t}$) that passes on the test suite. This indicates the model has not just overfitted into the natural language specification but can reason about executing the specified test and incorporate that into the implementation. 
\name measures the performance of a model $L$ in SR using $S_{SR}$ metric as below: 

\begin{small}
\vspace{-5pt}
\begin{equation}
\label{eq:SSR}
S_{SR}(L,S,t) = (1-Pass_{\langle C_S,T \rangle}) \times Pass_{\langle C_{S+t},T \rangle}
\end{equation}
\end{small}

$Pass_{\langle C_S,T \rangle}$ is $1$, if the test suite $T$ passes on $C_S$. Similarly, $Pass_{\langle C_{S+t},T \rangle}$ is $1$ if $T$ passes on $C_{S+t}$. 
Similar to the previous task, \name calculates the collective $R_{SR}$ values for a set of $m$ programs in benchmark $B_{|m|}$, considering the following two factors: A model that successfully generates more correct code by incorporating test data should be rewarded more. At the same time, the metric should avoid negative bias towards stronger models and challenging problems that cannot be solved, even with the hints from test data.

\begin{footnotesize}
\vspace{-8pt}
\begin{equation}
\label{eq:RSR}
   R_{SR}(L,B_{|m|})= Pass_{B_{|m|}} \times e^{(\dfrac{\sum\limits_{i=1}^{m} \llbracket S_{SR}(L,S_i \in B,t_i)=1 \rrbracket}{m})}
\end{equation}
\end{footnotesize}



In this equation, $Pass_{B_{|m|}} = \tfrac{\sum\limits_{i=1}^{m} Pass_{\langle C_{S_i},t_i \rangle}}{m}$ denotes the \emph{initial} success of LLM $L$, i.e., percentage of correct programs generated with only natural language specification. The model will be rewarded depending on the number of correct programs it can generate by reasoning about the test data in the specification (exponential growth rate to emphasize the change proportional to the previous state). By design, the equation takes a value between $0$ and $1$, making it a proper metric to compare the performance of LLMs with each other. 

\vspace{-5pt}
\subsection{\refactor}
\label{subsec:DSR-definition}
Ideally, LLMs should understand the overall code semantics, regardless of specific inputs and outputs, to analyze the code for different purposes and programming tasks. For example, in bug repair, while LLM is given one or multiple failing tests to localize and fix the bug, the ability to generalize the dynamic code semantics beyond the given test data will help the generated patch to pass on unseen tests. As a proxy to evaluate the general abilities of
LLMs in understanding code semantics, \name instructs LLMs to refactor code to a \emph{shorter}, \emph{semantically equivalent} version when possible. This requires reasoning about dynamic code semantics across all possible input/output pairs. We formally define this \emph{implicit} reasoning task as follows:

\vspace{3pt}
\textbf{Definition 3: Dynamic Semantics Reasoning (\CR).}
\label{subsec:CRDefination}
\label{subsec:ReasoningEval}
Given a program $C: C_I \rightarrow C_O$ and a test $t=\langle i \in C_I, o \in C_O\rangle$, LLM $L$ succeeds in \CR if it can refactor $C$ to $C^\prime$ such that $(\forall i \in C_I, C^{\prime}(i)=C(i)) \land (LoC(C^{\prime}) < LoC(C))$, where $LoC$ denotes lines of code. We argue that the objective of generating a shorter refactored code challenges LLMs more. Without that, LLMs may inject trivial/useless semantic-preserving or dead code to succeed. In the design of \CR, we make the following assumptions:

\begin{itemize}[leftmargin=*]
    \item Evaluating semantic equivalence is an NP-hard problem and one may be unable to validate semantic equivalence for the entire input space. Thereby, \name assumes the availability of test suite $T$ for $C$ and checks semantic equivalence considering the tests in $T$. 
    \item The goal of \name is to evaluate different aspects of LLMs' code reasoning capabilities specific to a given program. Without that, one cannot make a scientific conclusion about their abilities. Given that programs in the majority of benchmarks are standalone methods and are usually optimized, \name first refactor the original program $C$ through \emph{non-trivial} transformations (\S \ref{subsec:eval-DSR}) into $C^+$. It then asks LLMs to refactor $C^+$ and compares the generated code $C^\prime$ with the original $C$ program. 
\end{itemize}
 
Considering these assumptions \name measures the performance of a model $L$ in \CR using the following metric:

\vspace{-5pt}
\begin{footnotesize}
\begin{equation}
\label{eq:SCSR}
S_{\CR}(L, C, C^+, T) = Pass_{\langle C^{\prime},T \rangle} \times \frac{LoC(C) \times (1-\lfloor \dfrac{LoC(C^{\prime})}{LoC(C^+)} \rfloor)}{max(LoC(C^{\prime}), LoC(C))}
\end{equation}
\end{footnotesize}

\begin{figure*}
    \centering
    \includegraphics[width=0.98\linewidth]{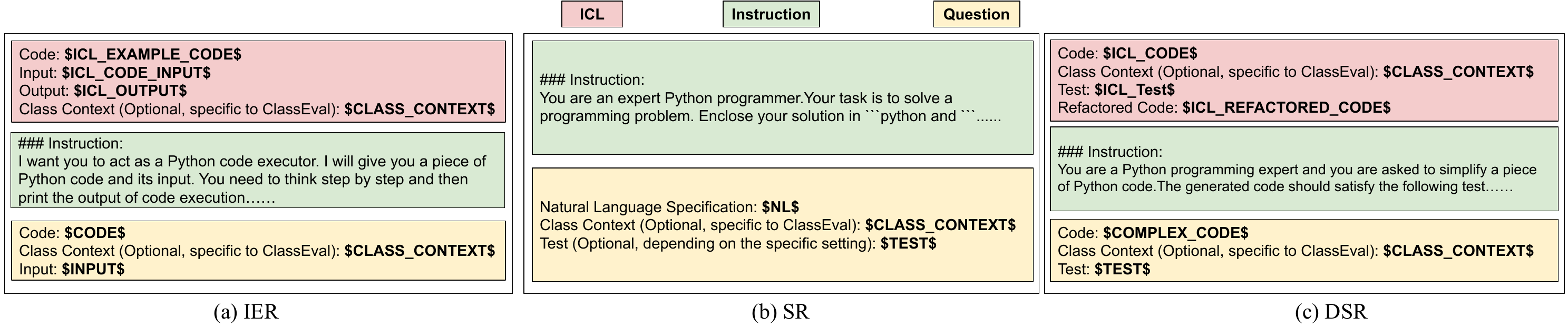}
    \vspace{-5pt}
    \caption{Prompt templates used for different reasoning tasks in \name}
    \label{fig:prompt-templates}
    \vspace{-10pt}
\end{figure*}

\begin{figure*}
    \centering
    \includegraphics[width=0.98\linewidth]{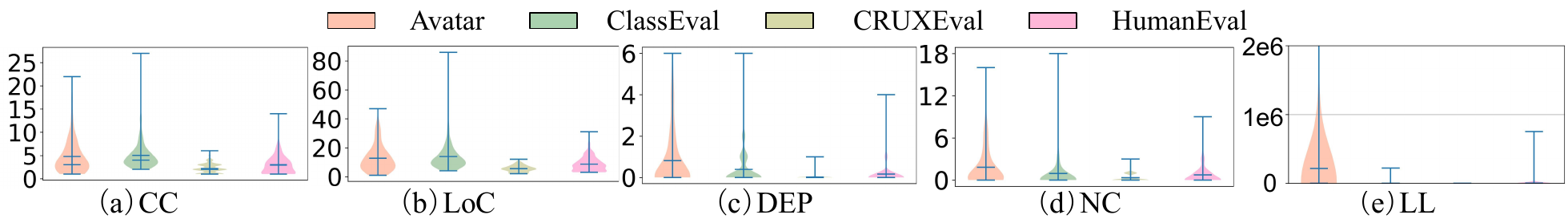}
    \vspace{-5pt}
    \caption{Distribution of the subject programs per different complexity metrics: Cyclomatic Complexity (CC), Lines of Code (LOC), Intra-class Dependencies (DEP), Nested Constructs (NC), and Loop Length (LL)}
    \label{fig:static-benchmark}
    \vspace{-10pt}
\end{figure*}

$Pass_{\langle C^{\prime},T \rangle}$ is $1$ if all the tests in $T$ pass on $C^{\prime}$, i.e., $C^{\prime}$ is semantically equivalent to $C^+$ and hence $C$. The closer the $LoC(C^{\prime})$ and $LoC(C)$ values, the better the model identifies and removes the code with no impact on semantics. While original programs are optimized in programming benchmarks, it is theoretically possible that LLM refactors $C^+$ to a code shorter than the original code. Hence, \name uses the maximum length of the generated and original code in the denominator. It also rules out the cases where LLM generates semantically equivalent programs longer than $C^+$ ($\lfloor\rfloor$ refers to the floor function), likely by adding useless or dead code. 
\name calculates the collective $R_{\CR}$ for the set of $m$ programs in benchmark $B_{|m|}$ as:

\vspace{-5pt}
\begin{small}
\begin{equation}
\label{eq:RCSR}
R_{\CR}(L,B_{|m|}) = \frac{\sum\limits_{i=1}^{m} S_{DSR}(L, C_i, C^+_i, T_i)}{m}
\end{equation}
\end{small}

\vspace{-5pt}
\subsection{Necessity of Reasoning Tasks} 
One can argue that some complex programming tasks, e.g., bug prediction or program repair, implicitly evaluate the code reasoning of the models. We strongly agree with this. At the same time, we argue that the achievements of LLMs in such tasks are not necessary due to their code understanding and code semantics reasoning. As we will show, there is no notable association between the success in code reasoning tasks and program repair (\S \ref{subsec:eval-APRandReasoning}). Our deep analysis shows that frontier reasoning LLMs, e.g., \claude, \omini, and \dsr, achieve the highest performance in both program repair and code reasoning, incorporating code reasoning in their problem-solving steps. 
Other models, however, can succeed in program repair by chance, hallucinations, or common patterns for fixing simple bugs (\S \ref{subsec:eval-APRandReasoning}). Our study highlights the limitations of LLMs in three fine-grained tasks carefully designed to evaluate their reasoning capabilities.

%% file: sections/3-Setups.tex
\vspace{-5pt}
\section{Experimental Setup}
\label{sec:setup}

\textbf{Subject LLMs.} We chose \major{$13$} pre-trained or instruction-tuned models, covering both general-purpose and Code LLMs. Limited by computing resources, we selected models with no larger than $34$B parameters that outperform the rest for programming tasks. Our subject LLMs are \gptf~\cite{chatgpt4}, \gemini~\cite{team2023gemini}, \codellama~(Instruct-13b, Base-13b, and Instruct-34b)~\cite{roziere2023code}, \deepseek~(Instruct-6.7b, Base-6.7b, and Instruct-33b)\cite{bi2024deepseek}, \sem(6.7b)\cite{ding2024semcoder}, and \starT~(15b)\cite{lozhkov2024starcoder}. 
We also evaluated \major{three recent reasoning LLMs: \claude~\cite{anthropic_claude_sonnet_4_6_system_card_2026}, \omini~\cite{o4-mini2025}, and \dsr~\cite{guo2025deepseek}}
We downloaded the open-access LLMs from HuggingFace~\cite{huggingface} and enforced temperature zero to ensure the reproducibility of results (more discussions in \S\ref{sec:threats}). For other parameters, we use the default setting of each model.

\textbf{Prompting Strategies.} 
Prompt crafting plays a crucial role in the performance of LLMs. Figure~\ref{fig:prompt-templates} illustrates the prompt templates used for various reasoning tasks, including:

\begin{itemize}[leftmargin=*]
    \item \emph{In-Context Example.} LLMs are instruction-tuned for code synthesis yet can learn new tasks through examples, i.e., In-Context Learning (ICL)~\cite{brown2020language,ye2023compositional,dong2022survey}. Since IER and DSR are new tasks, the prompt templates include ICL examples to introduce the task to LLMs. In addition to elaborating on how to perform the new task, the in-context example instructs the model for specific response formatting. \name's ICL examples are hand-crafted, reflecting our forecasted problem complexities. For example, in the task of IER, the code in the ICL example contains nested for loops and conditional statements, instructing LLMs how to reason step-by-step to solve the IER problem:
    \input{Listings/icl_example} 
    
    \item \emph{Instruction.} The next component is the instruction, where \name asks the model to solve the problem step by step in natural language (implicit Chain of Thought (CoT)). This step is necessary for the best performance for two reasons. First, LLMs are instruction-tuned through natural language instructions. Hence, they might understand tasks better in the presence of additional related natural language instructions. Second, CoT has been shown to improve the performance of the models in different tasks~\cite{wei2022chain,chen2022program}. 
    We chose CoT over Tree of Thought (ToT)~\cite{yao2024tree} and Graph of Thought (GoT)~\cite{besta2024graph} since their performance significantly depends on heuristics (rules or methods for selecting and guiding reasoning path selection). The design of heuristics in these techniques is problem-specific rather than task-specific, making their automated generation a separate research problem and out of the scope of this paper~\cite{shinn2023reflexion}. Given that \name focuses on comparing models and better understanding root causes, we anticipate improvement in prompt crafting results in the same conclusions.  
    
    \item \emph{Question.} The prompt template concludes with the main questions, i.e., asking the model to perform a specific reasoning task with the provided data. Depending on the problem and code, additional context will be provided in the \emph{Question} section. For example, we include the entire class context for \ceval programs, as there are intra-procedural dependencies between the methods, and the related context can be helpful for code reasoning. 
\end{itemize}

\name updates prompt templates per each program and adjusts them per each model, following the best prompting practices from official documents to ensure a fair evaluation. For example, \deepseek achieves the best performance by including the persona statement ``\emph{You are an AI programming assistant, utilizing the DeepSeek Coder model, developed by DeepSeek Company},'' as this sentence was used in their training phase. Upon receiving the response, \name automatically parses it and computes the metrics in Equations~\ref{eq:SIER}--\ref{eq:RCSR}. 

\textbf{Subject Programs.} We chose subject programs from widely used datasets: Avatar~\cite{ahmad2021avatar}, ClassEval~\cite{du2024evaluating}, CRUXEval~\cite{gu2024cruxeval}, and HumanEval~\cite{humaneval}. Although \name framework is programming language agnostic, all these programs are in Python, leaving us with $1450$ Python programs for evaluation (the column \emph{\#Subject} in Table~\ref{tab:IER}). 
We further evaluated the diversity of these programs in terms of cyclomatic complexity (CC), length of programs (LoC), intra-class dependency (DEP), existence of nested constructs (NC), and length of recursion (LL). Figure~\ref{fig:static-benchmark} compares the programs across datasets concerning these complexity metrics. 


The programs in \avatar and \ceval, on average, have higher Cyclomatic Complexity (CC)~\cite{gill1991cyclomatic} compared to \crux and \heval (Figure~\ref{fig:static-benchmark}-a). This means more independent execution paths within the programs in these benchmarks, potentially challenging LLMs to decide on the correct control flow path per given inputs. They are also longer in terms of the lines of code (Figure~\ref{fig:static-benchmark}-b), challenging the attention span of LLMs~\cite{liu2024lost}. Next, we measured the intra-class dependency (DEP) between the methods used to implement the programs. This is especially important since it challenges the ability to switch contexts from one method to another. While \ceval has more methods in the classes compared to \avatar, the DEP values for its program are smaller on average (Figure~\ref{fig:static-benchmark}-c). 
We also measured the number of nested constructs (NC), as reasoning about them is intuitively more challenging, even for humans (Figure~\ref{fig:static-benchmark}-d): programs in \avatar dataset, on average, have more nested constructs than other programs. Finally, we measure these programs' average Loop Lengths (LL). Again, this property is intuitively more challenging to reason about, as longer loops require memorizing more variable states and incorporating that into reasoning. To collect these numbers, we executed them through existing tests and measured the number of iterations per loop. Again, \avatar has more complex programs concerning this metric, i.e., there are programs with nested loops and lengths of over $2e^6$ iterations (Figure~\ref{fig:static-benchmark}-e). 
\input{Table/Table-IER}
We will use these complexity metrics to explain the observed results in the remainder of the paper. \crux, used in recent papers to evaluate LLMs on code execution reasoning, falls behind other benchmarks concerning different complexity metrics. Our experiments show that LLMs achieve the highest reasoning rate in our proposed tasks on \crux, which should raise concerns about using simple benchmarks and claiming victory on code reasoning for new LLMs. 


%% file: Listings/icl_example.tex
\begin{lstlisting}[style=stylewithcommentPy, columns=fullflexible]
@@Consider the following code:@@
def sum_of_integer(N, A, B):
   sum_1 = 0
   for i in range(1,N+1):
       sum_order = 0
       i_str = str(i)
       n = len(i_str)
       for j in range(0,n):
           sum_order += int(i_str[j])
       if A <= sum_order <= B:
           sum_1 += i
   return sum_1
@@[Question]@@
What |is| the |return| value of `sum_of_integer(20, 2, 5)?`
@@[Answer]@@
The variable N, variable A |and| variable B are initialized 
to 20, 2, |and| 5, respectively. Variable sum_1 |is| 
initialized to 0, which will be used to accumulate the  
|sum| of numbers meeting the condition. The code then  
enters a loop that iterates |from| 1 to N (inclusive), 
meaning it will consider numbers |from| 1 to 20. For each 
number i |in| this |range|, it calculates the |sum| of  its 
digits |and| stores it |in| sum_order. The code checks |if| 
sum_order |is| within the |range| [A, B], which |is| [2, 5] 
|in| this case. If it |is|, it adds the current number i to 
sum_1. The condition |is| met when  i |is| 2,3,4,5,11,12,13,14 
|and| 20. After the loop finishes, the code prints the 
final value of sum_1, which |is| 84.
@@[Output]@@
84
\end{lstlisting}

%% file: Table/Table-IER.tex
\begin{table*}[t]
\caption{\revision{Performance of subject LLMs in independent execution reasoning measured by $R_{IER}$ in Equation~\ref{RIER}. We highlight the top three best-performing models with {\color[HTML]{FE0000} red (\nth{1})}, {\color[HTML]{009901} green (\nth{2})}, and {\color[HTML]{3531FF} blue (\nth{3})}.}}
\vspace{-5pt}
\label{tab:IER}
\resizebox{\textwidth}{!}{
\begin{tabular}{|c|c|ccccccccccccc|}
\hline
 &
   &
  \multicolumn{13}{c|}{\textbf{Subject LLMs}} \\ \cline{3-15} 
 &
   &
  \multicolumn{3}{c|}{\textbf{CodeLlama}} &
  \multicolumn{3}{c|}{\textbf{DeepSeek-Coder}} &
  \multicolumn{1}{c|}{} &
  \multicolumn{1}{c|}{} &
  \multicolumn{1}{c|}{} &
  \multicolumn{1}{c|}{} &
  \multicolumn{1}{c|}{} &
  \multicolumn{1}{c|}{} &
   \\
\multirow{-3}{*}{\textbf{Dataset}} &
  \multirow{-3}{*}{\textbf{\#Subjects}} &
  \multicolumn{1}{c|}{\textbf{(Inst-13b)}} &
  \multicolumn{1}{c|}{\textbf{(Base-13b)}} &
  \multicolumn{1}{c|}{\textbf{(Inst-34b)}} &
  \multicolumn{1}{c|}{\textbf{(Inst-6.7b)}} &
  \multicolumn{1}{c|}{\textbf{(Base-6.7b)}} &
  \multicolumn{1}{c|}{\textbf{(Inst-33b)}} &
  \multicolumn{1}{c|}{\multirow{-2}{*}{\textbf{\begin{tabular}[c]{@{}c@{}}\sem\\ (6.7b)\end{tabular}}}} &
  \multicolumn{1}{c|}{\multirow{-2}{*}{\textbf{\begin{tabular}[c]{@{}c@{}}StarCoder2\\ (15b)\end{tabular}}}} &
  \multicolumn{1}{c|}{\multirow{-2}{*}{\textbf{\begin{tabular}[c]{@{}c@{}}Gemini-1.5-\\ Pro\end{tabular}}}} &
  \multicolumn{1}{c|}{\multirow{-2}{*}{\textbf{\begin{tabular}[c]{@{}c@{}}GPT-4-\\ Turbo\end{tabular}}}} &
  \multicolumn{1}{c|}{\multirow{-2}{*}{\textbf{\begin{tabular}[c]{@{}c@{}}\major{Claude-Sonnet-}\\ \major{4.6}\end{tabular}}}} &
  \multicolumn{1}{c|}{\multirow{-2}{*}{\textbf{\begin{tabular}[c]{@{}c@{}}\revision{DeepSeek-}\\ \revision{R1}\end{tabular}}}} &
  \multirow{-2}{*}{\textbf{\begin{tabular}[c]{@{}c@{}}\revision{o4-}\\ \revision{mini}\end{tabular}}} \\ \hline
\textbf{Avatar} &
  86 &
  \multicolumn{1}{c|}{23.26\%} &
  \multicolumn{1}{c|}{24.91\%} &
  \multicolumn{1}{c|}{24.42\%} &
  \multicolumn{1}{c|}{22.09\%} &
  \multicolumn{1}{c|}{25.58\%} &
  \multicolumn{1}{c|}{{41.86\%}} &
  \multicolumn{1}{c|}{30.23\%} &
  \multicolumn{1}{c|}{32.56\%} &
  \multicolumn{1}{c|}{{54.65\%}} &
  \multicolumn{1}{c|}{61.63\%} &
  \multicolumn{1}{c|}{{\color[HTML]{FE0000} 84.88\%}} &
  \multicolumn{1}{c|}{\color[HTML]{3531FF} 80.23\%} &
  \multicolumn{1}{c|}{\color[HTML]{009901} 83.72\%}
  \\ \hline
\textbf{ClassEval} &
  400 &
  \multicolumn{1}{c|}{43.50\%} &
  \multicolumn{1}{c|}{32.50\%} &
  \multicolumn{1}{c|}{47.25\%} &
  \multicolumn{1}{c|}{50.50\%} &
  \multicolumn{1}{c|}{50.00\%} &
  \multicolumn{1}{c|}{{65.00\%}} &
  \multicolumn{1}{c|}{61.00\%} &
  \multicolumn{1}{c|}{55.25\%} &
  \multicolumn{1}{c|}{{74.00\%}} &
  \multicolumn{1}{c|}{ 77.50\%} &
  \multicolumn{1}{c|}{{\color[HTML]{FE0000} 92.50\%}} &
   \multicolumn{1}{c|}{\color[HTML]{3531FF} 88.25\%} &
  \multicolumn{1}{c|}{\color[HTML]{009901} 91.50\%}
  \\ \hline
\textbf{CRUXEval} &
  800 &
  \multicolumn{1}{c|}{41.50\%} &
  \multicolumn{1}{c|}{37.75\%} &
  \multicolumn{1}{c|}{48.13\%} &
  \multicolumn{1}{c|}{42.38\%} &
  \multicolumn{1}{c|}{41.50\%} &
  \multicolumn{1}{c|}{{58.88\%}} &
  \multicolumn{1}{c|}{52.00\%} &
  \multicolumn{1}{c|}{53.00\%} &
  \multicolumn{1}{c|}{{74.44\%}} &
  \multicolumn{1}{c|}{82.63\%} &
  \multicolumn{1}{c|}{{\color[HTML]{FE0000} 98.50\%}} &
   \multicolumn{1}{c|}{\color[HTML]{3531FF} 96.93\%} &
  \multicolumn{1}{c|}{\color[HTML]{009901} 98.25\%}
  \\ \hline
\textbf{HumanEval} &
  164 &
  \multicolumn{1}{c|}{50.00\%} &
  \multicolumn{1}{c|}{44.44\%} &
  \multicolumn{1}{c|}{55.56\%} &
  \multicolumn{1}{c|}{61.11\%} &
  \multicolumn{1}{c|}{53.70\%} &
  \multicolumn{1}{c|}{{64.81\%}} &
  \multicolumn{1}{c|}{{64.81\%}} &
  \multicolumn{1}{c|}{46.40\%} &
  \multicolumn{1}{c|}{{83.33\%}} &
  \multicolumn{1}{c|}{ 93.21\%} &
  \multicolumn{1}{c|}{\color[HTML]{FE0000}{97.56\%}} &
  \multicolumn{1}{c|}{\color[HTML]{FE0000} 97.56\%} &
  \multicolumn{1}{c|}{\color[HTML]{3531FF} 96.95\%}
  \\ \hline

\textbf{Total} &
  1450 &
  \multicolumn{1}{c|}{41.93\%} &
  \multicolumn{1}{c|}{36.30\%} &
  \multicolumn{1}{c|}{47.32\%} &
  \multicolumn{1}{c|}{45.54\%} &
  \multicolumn{1}{c|}{44.28\%} &
  \multicolumn{1}{c|}{{60.23\%}} &
  \multicolumn{1}{c|}{55.19\%} &
  \multicolumn{1}{c|}{51.11\%} &
  \multicolumn{1}{c|}{{74.15\%}} &
  \multicolumn{1}{c|}{ 81.17\%} &
  \multicolumn{1}{c|}{\color[HTML]{FE0000}{95.93\%}} &
  \multicolumn{1}{c|}{\color[HTML]{3531FF} 90.43\%} &
  \multicolumn{1}{c|}{\color[HTML]{009901} 92.52\%}
  \\ \hline
 
\rowcolor{lightgray} \multicolumn{2}{|c|}{\textbf{$\rho_{CC}$}} &
  \multicolumn{1}{c|}{-0.38} &
  \multicolumn{1}{c|}{-0.43} &
  \multicolumn{1}{c|}{-0.89} &
  \multicolumn{1}{c|}{-0.47} &
  \multicolumn{1}{c|}{-0.39} &
  \multicolumn{1}{c|}{-0.42} &
  \multicolumn{1}{c|}{-0.41} &
  \multicolumn{1}{c|}{-0.63} &
  \multicolumn{1}{c|}{-0.61} &
   \multicolumn{1}{c|}{-0.68} &
   \multicolumn{1}{c|}{-0.47} &
  \multicolumn{1}{c|}{-0.54} &
  \multicolumn{1}{c|}{-0.52}
  \\ \hline

\rowcolor{lightgray} \multicolumn{2}{|c|}{\textbf{$\rho_{LoC}$}} &
  \multicolumn{1}{c|}{-0.40} &
  \multicolumn{1}{c|}{-0.47} &
  \multicolumn{1}{c|}{-0.55} &
  \multicolumn{1}{c|}{-0.57} &
  \multicolumn{1}{c|}{-0.52} &
  \multicolumn{1}{c|}{-0.32} &
  \multicolumn{1}{c|}{-0.53} &
  \multicolumn{1}{c|}{-0.49} &
  \multicolumn{1}{c|}{-0.31} &
  \multicolumn{1}{c|}{-0.50} &
  \multicolumn{1}{c|}{-0.58} &
 \multicolumn{1}{c|}{-0.44} &
  \multicolumn{1}{c|}{-0.47} 
  \\ \hline

\rowcolor{lightgray}   \multicolumn{2}{|c|}{\textbf{$\rho_{DEP}$}} &
  \multicolumn{1}{c|}{-0.51} &
  \multicolumn{1}{c|}{-0.38} &
  \multicolumn{1}{c|}{-0.57} &
  \multicolumn{1}{c|}{-0.40} &
  \multicolumn{1}{c|}{-0.29} &
  \multicolumn{1}{c|}{-0.68} &
  \multicolumn{1}{c|}{-0.82} &
  \multicolumn{1}{c|}{-0.72} &
  \multicolumn{1}{c|}{-0.54} &
  \multicolumn{1}{c|}{-0.68} &
  \multicolumn{1}{c|}{-0.53} &
  \multicolumn{1}{c|}{-0.66} &
  \multicolumn{1}{c|}{-0.61}
  \\ \hline

\rowcolor{lightgray}   \multicolumn{2}{|c|}{\textbf{$\rho_{NC}$}} &
  \multicolumn{1}{c|}{-0.36} &
  \multicolumn{1}{c|}{-0.21} &
  \multicolumn{1}{c|}{-0.13} &
  \multicolumn{1}{c|}{-0.51} &
  \multicolumn{1}{c|}{-0.42} &
  \multicolumn{1}{c|}{-0.49} &
  \multicolumn{1}{c|}{-0.54} &
  \multicolumn{1}{c|}{-0.35} &
  \multicolumn{1}{c|}{-0.49} &
  \multicolumn{1}{c|}{-0.61} &
  \multicolumn{1}{c|}{-0.55} &
  \multicolumn{1}{c|}{-0.52} &
  \multicolumn{1}{c|}{-0.51}
  \\ \hline

\rowcolor{lightgray}   \multicolumn{2}{|c|}{\textbf{$\rho_{LL}$}} &
  \multicolumn{1}{c|}{-0.36} &
  \multicolumn{1}{c|}{-0.49} &
  \multicolumn{1}{c|}{-0.39} &
  \multicolumn{1}{c|}{-0.51} &
  \multicolumn{1}{c|}{-0.29} &
  \multicolumn{1}{c|}{-0.56} &
  \multicolumn{1}{c|}{-0.71} &
  \multicolumn{1}{c|}{-0.59} &
  \multicolumn{1}{c|}{-0.22} &
  \multicolumn{1}{c|}{-0.35} &
  \multicolumn{1}{c|}{-0.41} &
 \multicolumn{1}{c|}{-0.40} &
  \multicolumn{1}{c|}{-0.32}
  \\ \hline
\end{tabular}
}
\vspace{-10pt}
\end{table*}

%% file: sections/4-Evaluation.tex
\vspace{-5pt}
\section{Empirical Evaluation}

We leverage \name to investigate how well LLMs reason about subject programs (\S\ref{subsec:eval-IER}--\S\ref{subsec:eval-DSR}). We further perform an in-depth analysis of the reasoning failures to understand the challenging factors (\S\ref{subsec:eval-factors}) and demonstrate the necessity of using proposed reasoning tasks to evaluate LLMs (\S\ref{subsec:eval-different-reasoning}--\S\ref{subsec:agentic}). Finally, we compare the performance of \name with an existing code reasoning tool, \reval, for the common task of output prediction (\S\ref{subsec:comparison}). 

\input{sections/4-1-IER}

\input{sections/4-2-SR}

\input{sections/4-3-DSR}

\input{sections/4-4-Factors}

\input{sections/4-5-Different-Reasoning-Tasks}
\input{sections/4-6-APR-Comparison}
\input{sections/4-7-Agentic-System}
\input{sections/4-8-REval-Comparison}

%% file: sections/4-1-IER.tex
\vspace{-5pt}
\subsection{RQ1: Performance of LLMs in \IER}
\label{subsec:eval-IER}

To evaluate the performance of LLMs on \IER, \name prompts the models using the prompt template shown in Figure~\ref{fig:prompt-templates}-a. Table~\ref{tab:IER} shows the result of this experiment\footnote{Note that our results for \crux might be different from the numbers reported in their paper because (1) we consider the temperature $0$ for our experiments and (2) our prompt template is different.}.
\revision{Overall, \textbf{the frontier, reasoning models outperform the best non-reasoning API-access model (\gptf) with large margins of \major{$14.76\%$ (\claude)}, $11.35\%$ (\omini), and $9.26\%$ (\dsr)}. They also outperform the best open-access model (\deepseek-Inst-33b) by $35.70\%$, $32.29\%$, and $30.20\%$, respectively.} We speculate the size of such models (in terms of the number of parameters) plays an important role when compared to smaller models. Furthermore, these models are reinforcement-learned and instruction-tuned with high-quality and large-scale human feedback, making them follow instructions better and outperform \IER.
\textbf{Within the family of models, LLMs with more parameters always outperform smaller ones on \IER}: the $R_{IER}$ improves from $41.93\%$ (\codellama-Instruct-13b) to $47.32\%$ (\codellama-Instruct-34b) and from $45.54\%$ (\deepseek-Instruct-6.7b) to $60.23\%$ (\deepseek-Instruct-33b) \revision{and $90.43\%$ (\dsr)}.

\textbf{Instruction-tuning improves the performance of LLMs in \IER}: for \codellama-13b, and \deepseek-6.7b, the instruction-tuned version outperforms the base with the margins of $5.63\%$, and $1.26\%$, respectively, mainly because the instruction-tuned LLMs follow prompt instructions better. For \sem(6.7b), fine-tuned on \deepseek-Base-6.7b with \emph{execution data} \major{(natural language reasoning on how the input of a piece of code evolves to the output synthesized by a more powerful LLM)}, the improvement is $10.91\%$. \textbf{\sem also outperforms instruction-tuned models of the same size or even larger, demonstrating the impact of execution-aware fine-tuning in better code reasoning.}

LLMs struggle to reason about programs in \avatar more than other benchmarks. As discussed before (\S \ref{sec:setup}), \textbf{programs in \avatar have more complex code constructs and semantics (\S\ref{sec:setup}), challenging LLMs to track how the inputs turn into output through code execution}. 
Furthermore, prior research has shown that LLMs overfit into widely used benchmarks and do not generalize well beyond them~\cite{jain2024livecodebench}, which may explain why the performance of LLMs in \heval is higher than other benchmarks.

%% file: sections/4-2-SR.tex
\vspace{-8pt}
\subsection{RQ2: Performance of LLMs in \SR}
\label{subsec:eval-SR}

\input{Table/Table-SR}

\begin{figure*}[t]
    \centering
    \includegraphics[width=0.98\linewidth]{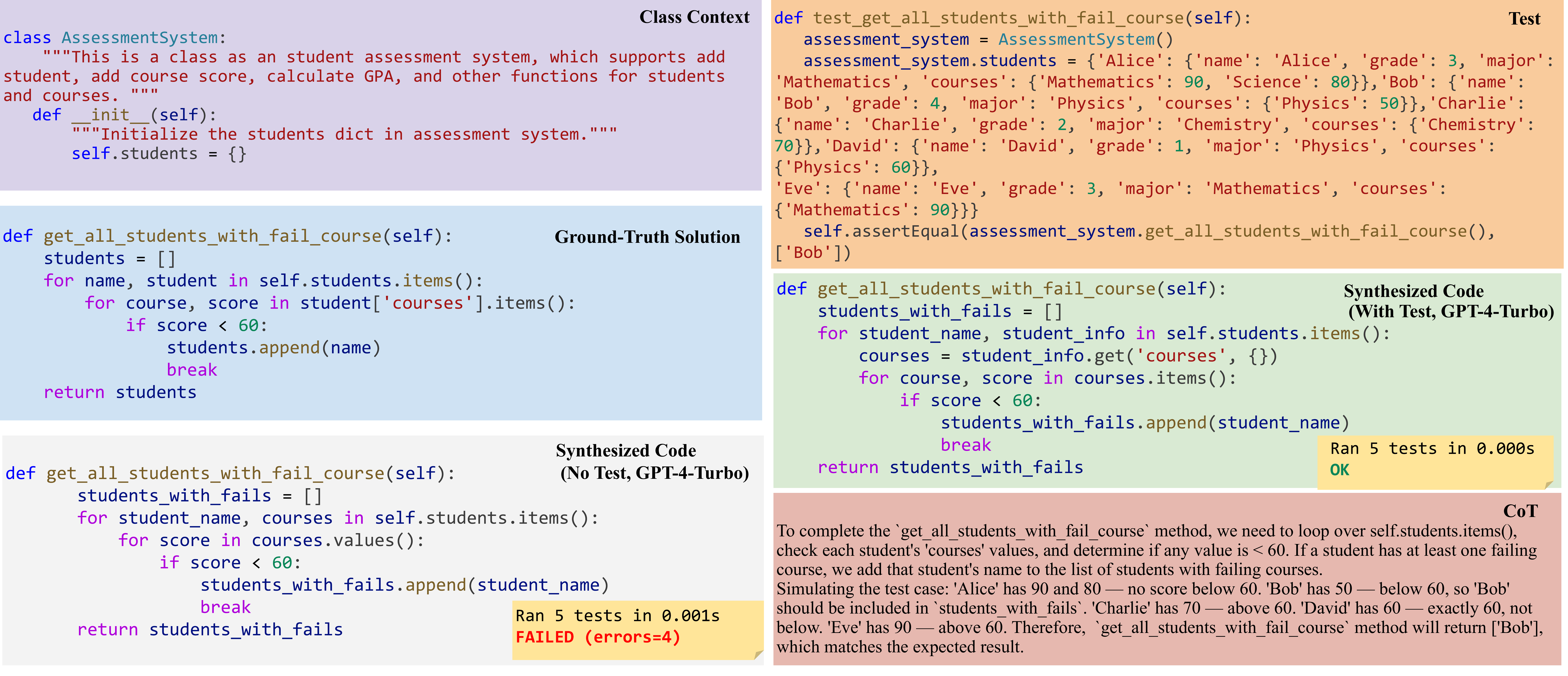}
    \caption{\revision{Performance of \gptf in code synthesis under \emph{No Test} and \emph{With Test} settings or SR task for program ClassEval\_4}}
    \vspace{-10pt}
    \label{fig:sr-tests}
\end{figure*}

To evaluate the abilities of LLMs on \SR, \name prompts LLMs for code synthesis under the following two settings, using the prompt template in Figure~\ref{fig:prompt-templates}-b:
\emph{(1) Natural language specification only (\textbf{No Test}).} \name uses only the natural language specification to prompt the model for code synthesis. It validates the generated code using all the existing ground-truth tests. This setting serves as the baseline and mimics how users typically prompt LLMs for code synthesis.
\emph{(2) Natural language specification plus one ground-truth input-output (\textbf{With Test})}. Under this setting, \name randomly selects a ground-truth test and adds it to the specification. It validates the synthesized code using \emph{all the existing tests}. 

We use \heval and \ceval for this experiment, as the other two datasets do not have natural language specifications for prompting the models. The results in Table \ref{tab:sr} show that \textbf{the performance of LLMs in code synthesis with test data included in the specification, i.e., measured by pass@1, improves by \revision{$5.36\%$} on average}\footnote{Note that our numbers may not precisely match the leaderboards', as we used the temperature $0$ for our experiments.}. The improvement is \emph{higher} on \ceval (\revision{$9.78\%$}) compared with \heval (\revision{$0.94\%$}), although the average success under \emph{With Test} setting in \ceval is \emph{lower} than \heval. Based on our in-depth investigation of the \ceval cases that LLMs failed under \emph{No Test} but succeeded under \emph{With Test} settings, we speculate this is due to the \textbf{ambiguous natural language specifications} in this dataset compared to HumanEval. 

\major{To mitigate potential sensitivity to input sampling, we repeat the experiment with a set of previously unused inputs (i.e., inputs not included in Table~\ref{tab:sr}). The results are consistent across two runs: on average, including the test data in the specification improves the pass@1 by $5.56\%$ across two runs, and the improvement on \ceval ($9.68\%$ on average) is higher than \heval ($1.44\%$ on average).}

In the example of Figure~\ref{fig:sr-tests} from ClassEval\_4, the natural language specification is \texttt{\footnotesize{Get all students who have any score below 60}}. It also identifies the output as \texttt{\footnotesize{list of str, student names}}. The provided class context (purple box on the top left) includes the class declaration, description, and constructor\footnote{The typos in the class description are part of the dataset, not our mistake.}. The natural language specification and provided context in the benchmark are ambiguous and incomplete; thereby, \gptf fails to synthesize a correct code: running the tests on the code generated under \emph{No Test} (gray box on bottom left) setting results in a \texttt{\small{Type Error}} due to comparing a string value with an integer (\texttt{\small{if score < 60}}). Including the test data (orange box on the top right) provides more information about the student information structure, helping LLMs synthesize a code that passes all the tests. 
\revision{From the CoT (red box on bottom right), we observe that \gptf reasons step by step about code constructs in the synthesis,  while simulating the execution of the test case to verify its correctness. This example illustrates how the model integrates its understanding of execution output and logical control flow, which is beyond merely hard constraints to satisfy, with code synthesis, to produce correct (test-passing) code.}

These results show that LLMs can incorporate the test data into the code synthesis process, although to a limited extent. \textbf{When the natural language is ambiguous and the relevant context is incomplete, including the test data is more helpful for models synthesizing correct code. When the performance of the models is close under the \emph{No Test} setting, models with better SR reasoning, i.e., those that can incorporate test data into generating a correct code, will be rewarded more.} For example, \gptf and \gemini achieve $61.46\%$ and $60.49\%$ success rate under the \emph{No Test} setting in \ceval. \gemini succeed in SR for more cases compared to \gptf, resulting in the $R_{SR}$ value of $70.99\%$ compared to $66.13\%$ of \gptf. 

%% file: Table/Table-SR.tex
\begin{table*}[]
\centering
\caption{\revision{Performance of subject LLMs in specification reasoning measured by $R_{SR}$ in Equation 4 and detailed results on code synthesis under different prompt settings (demonstrated by pass@1). The $\greenuparrow$ symbol indicates the improvement from \textit{No Test} to \textit{With Test}. We highlight the top three best-performing models in terms of $R_{SR}$ with {\color[HTML]{FE0000} red (\nth{1})}, {\color[HTML]{009901} green (\nth{2})}, and {\color[HTML]{3531FF} blue (\nth{3})}}.}
\vspace{-5pt}
\label{tab:sr}
\resizebox{\textwidth}{!}{
\begin{tabular}{|c|c|ccccccccccccc|}
\hline
 &
   &
  \multicolumn{13}{c|}{\textbf{Subject LLMs}} \\ \cline{3-15} 
 &
   &
  \multicolumn{3}{c|}{\textbf{CodeLlama}} &
  \multicolumn{3}{c|}{\textbf{DeepSeek-Coder}} &
  \multicolumn{1}{c|}{} &
  \multicolumn{1}{c|}{} &
   \multicolumn{1}{c|}{} &
  \multicolumn{1}{c|}{} &
  \multicolumn{1}{c|}{} &
  \multicolumn{1}{c|}{} &
   \\ 
\multirow{-3}{*}{\textbf{Dataset}} &
  \multirow{-3}{*}{\textbf{Settings}} &
  \multicolumn{1}{c|}{\textbf{(Inst-13b)}} &
  \multicolumn{1}{c|}{\textbf{(Base-13b)}} &
  \multicolumn{1}{c|}{\textbf{(Inst-34b)}} &
  \multicolumn{1}{c|}{\textbf{(Inst-6.7b)}} &
  \multicolumn{1}{c|}{\textbf{(Base-6.7b)}} &
  \multicolumn{1}{c|}{\textbf{(Inst-33b)}} &
  \multicolumn{1}{c|}{\multirow{-2}{*}{\textbf{\begin{tabular}[c]{@{}c@{}}\sem\\ (6.7b)\end{tabular}}}} &
  \multicolumn{1}{c|}{\multirow{-2}{*}{\textbf{\begin{tabular}[c]{@{}c@{}}StarCoder2\\ (15b)\end{tabular}}}} &
  \multicolumn{1}{c|}{\multirow{-2}{*}{\textbf{\begin{tabular}[c]{@{}c@{}}Gemini-1.5-\\ Pro\end{tabular}}}} &
  \multicolumn{1}{c|}{\multirow{-2}{*}{\textbf{\begin{tabular}[c]{@{}c@{}}GPT-4-\\ Turbo\end{tabular}}}}&
  \multicolumn{1}{c|}{\multirow{-2}{*}{\textbf{\begin{tabular}[c]{@{}c@{}}\major{Claude-Sonnet-}\\ \major{4.6}\end{tabular}}}} &
 \multicolumn{1}{c|}{\multirow{-2}{*}{\textbf{\begin{tabular}[c]{@{}c@{}}\revision{DeepSeek-}\\ \revision{R1}\end{tabular}}}} &
  \multicolumn{1}{c|}{\multirow{-2}{*}{\textbf{\begin{tabular}[c]{@{}c@{}}\revision{o4-}\\ \revision{mini}\end{tabular}}}}
  \\ \hline
 &
  \textbf{No Test} &
  \multicolumn{1}{c|}{46.34\%} &
  \multicolumn{1}{c|}{31.10\%} &
  \multicolumn{1}{c|}{46.34\%} &
  \multicolumn{1}{c|}{76.83\%} &
  \multicolumn{1}{c|}{50.61\%} &
  \multicolumn{1}{c|}{71.34\%} &
  \multicolumn{1}{c|}{76.83\%} &
  \multicolumn{1}{c|}{46.34\%} &
  \multicolumn{1}{c|}{81.10\%} &
  \multicolumn{1}{c|}{89.63\%} & 
  \multicolumn{1}{c|}{96.95\%} & 
  \multicolumn{1}{c|}{93.29\%} &
  \multicolumn{1}{c|}{96.95\%}
   \\ \cline{2-15} 
   &
  \textbf{With Test} &
  \multicolumn{1}{c|}{\textcolor{black}{48.17\%}  $\greenuparrow$} &
  \multicolumn{1}{c|}{29.88\%} &
  \multicolumn{1}{c|}{\textcolor{black}{47.56\%}  $\greenuparrow$} &
  \multicolumn{1}{c|}{76.83\%} &
  \multicolumn{1}{c|}{48.17\%} &
  \multicolumn{1}{c|}{\textcolor{black}{76.83\%} $\greenuparrow$} &
  \multicolumn{1}{c|}{75.00\%} &
  \multicolumn{1}{c|}{\textcolor{black}{49.39\%} $\greenuparrow$} &
  \multicolumn{1}{c|}{\textcolor{black}{83.54\%} $\greenuparrow$} &
  \multicolumn{1}{c|}{\textcolor{black}{90.24\%} $\greenuparrow$} &
  \multicolumn{1}{c|}{96.95\%} & 
  \multicolumn{1}{c|}{\textcolor{black}{96.34\%} $\greenuparrow$} &
  \multicolumn{1}{c|}{96.95\%}
  \\ \cline{2-15}

\multirow{-3}{*}{\textbf{HumanEval}} &
  \textbf{$R_{SR}$} &
  \multicolumn{1}{c|}{47.49\%} &
  \multicolumn{1}{c|}{32.29\%} &
  \multicolumn{1}{c|}{47.78\%} &
  \multicolumn{1}{c|}{77.91\%} &
  \multicolumn{1}{c|}{53.45\%} &
  \multicolumn{1}{c|}{78.17\%} &
  \multicolumn{1}{c|}{$78.39\%$} &
  \multicolumn{1}{c|}{48.66\%} &
  \multicolumn{1}{c|}{ $85.79\%$} &
  \multicolumn{1}{c|}{$92.97\%$} &
  \multicolumn{1}{c|}{\color[HTML]{009901} 96.95\%} & 
  \multicolumn{1}{c|}{ \color[HTML]{FE0000} 97.36\%} &
  \multicolumn{1}{c|}{ \color[HTML]{009901} 96.95\%}
  \\ \hline

  \hline
 &
  \textbf{No Test} &
  \multicolumn{1}{c|}{42.86\%} &
  \multicolumn{1}{c|}{25.85\%} &
  \multicolumn{1}{c|}{45.37\%} &
  \multicolumn{1}{c|}{57.80\%} &
  \multicolumn{1}{c|}{42.93\%} &
  \multicolumn{1}{c|}{51.95\%} &
  \multicolumn{1}{c|}{41.71\%} &
  \multicolumn{1}{c|}{34.39\%} &
  \multicolumn{1}{c|}{60.49\%} &
  \multicolumn{1}{c|}{61.46\%} &
  \multicolumn{1}{c|}{78.29\%} & 
  \multicolumn{1}{c|}{ 72.44\%} &
  \multicolumn{1}{c|}{ 70.98\%}   
   \\ \cline{2-15} 
   &
  \textbf{With Test} &
  \multicolumn{1}{c|}{\textcolor{black}{48.29\%} $\greenuparrow$} &
  \multicolumn{1}{c|}{\textcolor{black}{42.20\%} $\greenuparrow$} &
  \multicolumn{1}{c|}{\textcolor{black}{51.71\%} $\greenuparrow$} &
  \multicolumn{1}{c|}{\textcolor{black}{61.46\%} $\greenuparrow$} &
  \multicolumn{1}{c|}{\textcolor{black}{46.59\%} $\greenuparrow$} &
  \multicolumn{1}{c|}{\textcolor{black}{62.93\%} $\greenuparrow$} &
  \multicolumn{1}{c|}{\textcolor{black}{47.07\%} $\greenuparrow$} &
  \multicolumn{1}{c|}{\textcolor{black}{42.68\%} $\greenuparrow$} &
  \multicolumn{1}{c|}{\textcolor{black}{72.20\%} $\greenuparrow$} &
  \multicolumn{1}{c|}{\textcolor{black}{69.76\%} $\greenuparrow$} &
  \multicolumn{1}{c|}{94.15\% $\greenuparrow$} & 
  \multicolumn{1}{c|}{ \textcolor{black}{90.98\%} $\greenuparrow$} &
  \multicolumn{1}{c|}{ \textcolor{black}{88.78\%} $\greenuparrow$}
  \\ \cline{2-15}

 
\multirow{-3}{*}{\textbf{ClassEval}} &
  \textbf{$R_{SR}$} &
  \multicolumn{1}{c|}{45.29\%} &
  \multicolumn{1}{c|}{30.59\%} &
  \multicolumn{1}{c|}{48.83\%} &
  \multicolumn{1}{c|}{$60.99\%$} &
  \multicolumn{1}{c|}{44.78\%} &
  \multicolumn{1}{c|}{ 61.47\%} &
  \multicolumn{1}{c|}{44.92\%} &
  \multicolumn{1}{c|}{36.64\%} &
  \multicolumn{1}{c|}{ $70.99\%$} &
 \multicolumn{1}{c|} { $66.13\%$} &
 \multicolumn{1}{c|}{\color[HTML]{FE0000} 92.64\%} & 
  \multicolumn{1}{c|}{ \color[HTML]{3531FF} 88.48\%} &
  \multicolumn{1}{c|}{ \color[HTML]{009901} 88.61\%}
  
  \\ \hline

  \rowcolor{lightgray} \multicolumn{2}{|c|}{\textbf{$\rho_{CC}$}} &
  \multicolumn{1}{c|}{-0.55} &
  \multicolumn{1}{c|}{-0.53} &
  \multicolumn{1}{c|}{-0.87} &
  \multicolumn{1}{c|}{-0.75} &
  \multicolumn{1}{c|}{-0.70} &
  \multicolumn{1}{c|}{-0.83} &
  \multicolumn{1}{c|}{-0.68} &
  \multicolumn{1}{c|}{-0.60} &
  \multicolumn{1}{c|}{-0.82} &
  \multicolumn{1}{c|}{-0.78} &
  \multicolumn{1}{c|}{-0.69\%} & 
  \multicolumn{1}{c|}{-0.71} &
  \multicolumn{1}{c|}{-0.69} 
  \\ \hline

\rowcolor{lightgray} \multicolumn{2}{|c|}{\textbf{$\rho_{LoC}$}} &
  \multicolumn{1}{c|}{-0.51} &
  \multicolumn{1}{c|}{-0.27} &
  \multicolumn{1}{c|}{-0.59} &
  \multicolumn{1}{c|}{-0.38} &
  \multicolumn{1}{c|}{-0.51} &
  \multicolumn{1}{c|}{-0.52} &
  \multicolumn{1}{c|}{-0.58} &
  \multicolumn{1}{c|}{-0.13} &
  \multicolumn{1}{c|}{-0.54} &
  \multicolumn{1}{c|}{-0.56} &
  \multicolumn{1}{c|}{-0.68\%} & 
\multicolumn{1}{c|}{-0.52} &
  \multicolumn{1}{c|}{-0.55} 
  \\ \hline

\rowcolor{lightgray}   \multicolumn{2}{|c|}{\textbf{$\rho_{DEP}$}} &
  \multicolumn{1}{c|}{-0.86} &
  \multicolumn{1}{c|}{-0.73} &
  \multicolumn{1}{c|}{-0.93} &
  \multicolumn{1}{c|}{-0.74} &
  \multicolumn{1}{c|}{-0.77} &
  \multicolumn{1}{c|}{-0.84} &
  \multicolumn{1}{c|}{-0.85} &
  \multicolumn{1}{c|}{-0.76} &
  \multicolumn{1}{c|}{-0.86} &
  \multicolumn{1}{c|}{-0.86} &
  \multicolumn{1}{c|}{-0.85\%} & 
\multicolumn{1}{c|}{-0.88} &
  \multicolumn{1}{c|}{-0.84} 
  \\ \hline

\rowcolor{lightgray}   \multicolumn{2}{|c|}{\textbf{$\rho_{NC}$}} &
  \multicolumn{1}{c|}{-0.71} &
  \multicolumn{1}{c|}{-0.82} &
  \multicolumn{1}{c|}{-0.90} &
  \multicolumn{1}{c|}{-0.81} &
  \multicolumn{1}{c|}{-0.74} &
  \multicolumn{1}{c|}{-0.71} &
  \multicolumn{1}{c|}{-0.62} &
  \multicolumn{1}{c|}{-0.43} &
  \multicolumn{1}{c|}{-0.77} &
  \multicolumn{1}{c|}{-0.84} &
  \multicolumn{1}{c|}{-0.67\%} & 
 \multicolumn{1}{c|}{-0.78} &
  \multicolumn{1}{c|}{-0.77} 
  \\ \hline

\rowcolor{lightgray}   \multicolumn{2}{|c|}{\textbf{$\rho_{LL}$}} &
  \multicolumn{1}{c|}{-0.41} &
  \multicolumn{1}{c|}{-0.40} &
  \multicolumn{1}{c|}{-0.29} &
  \multicolumn{1}{c|}{-0.47} &
  \multicolumn{1}{c|}{-0.29} &
  \multicolumn{1}{c|}{-0.52} &
  \multicolumn{1}{c|}{-0.34} &
  \multicolumn{1}{c|}{-0.12} &
  \multicolumn{1}{c|}{-0.50} &
  \multicolumn{1}{c|}{-0.53} &
  \multicolumn{1}{c|}{-0.39\%} & 
\multicolumn{1}{c|}{-0.43} &
  \multicolumn{1}{c|}{-0.46} 
  \\ \hline

\end{tabular}
}
\vspace{-10pt}
\end{table*}

%% file: sections/4-3-DSR.tex
\vspace{-5pt}
\subsection{RQ3: Performance of LLMs in \CR}
\label{subsec:eval-DSR}

\input{Table/Table-CR}

\begin{figure*}[]
    \centering
    \vspace{-8pt}
    \includegraphics[width=0.98\linewidth]{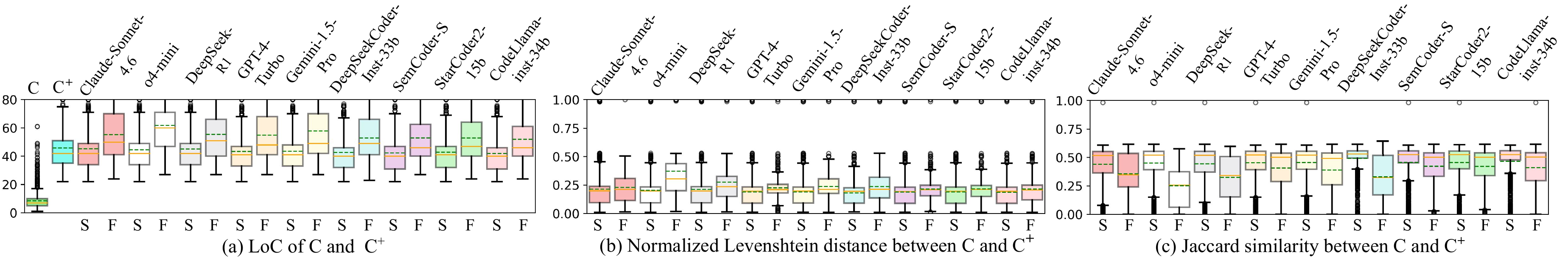}
    \vspace{-3pt}
    \caption{\major{
    Comparison between the size (a), Levenshtein distance (b), and Jaccard similarity (c) of $C$ and $C^+$ programs 
    }
    }
    \vspace{-10pt}
    \label{fig:dsr}
\end{figure*}

To evaluate the performance on \CR, \name prompts the models using the template shown in Figure~\ref{fig:prompt-templates}-c. \revision{We use a fixed ICL example containing only six transformations to convey the task’s high-level idea without disclosing excessive details of the code transformations.} To generate the programs required for proper evaluation of LLMs under this task (assumption two in \S \ref{subsec:DSR-definition}), \name implements and applies $20$ non-trivial, semantically-preserving transformations categorized into four groups: (1) creating more complex code structure by increasing the nested level of conditional (e.g., if blocks) and recursive structures (e.g., for and while loops), as well as introducing extra code constructs (e,g., try-except clauses and threads) into the program; (2) introducing widely used third-party APIs, e.g., \texttt{\small{base64}}, \texttt{\small{crypto}}, \texttt{\small{dateutil}}, \texttt{\small{numpy}}, \texttt{\small{scipy}}, and \texttt{\small{sklearn}}; (3) introducing inter/intra-procedural dependencies to code; and (4) renaming variables and functions. We list all the transformations in Table~\ref{tab:transformation}. All the transformations in this study are available on \name's artifact website~\cite{website}.

\begin{figure}[]
    \centering
    \includegraphics[width=\linewidth]{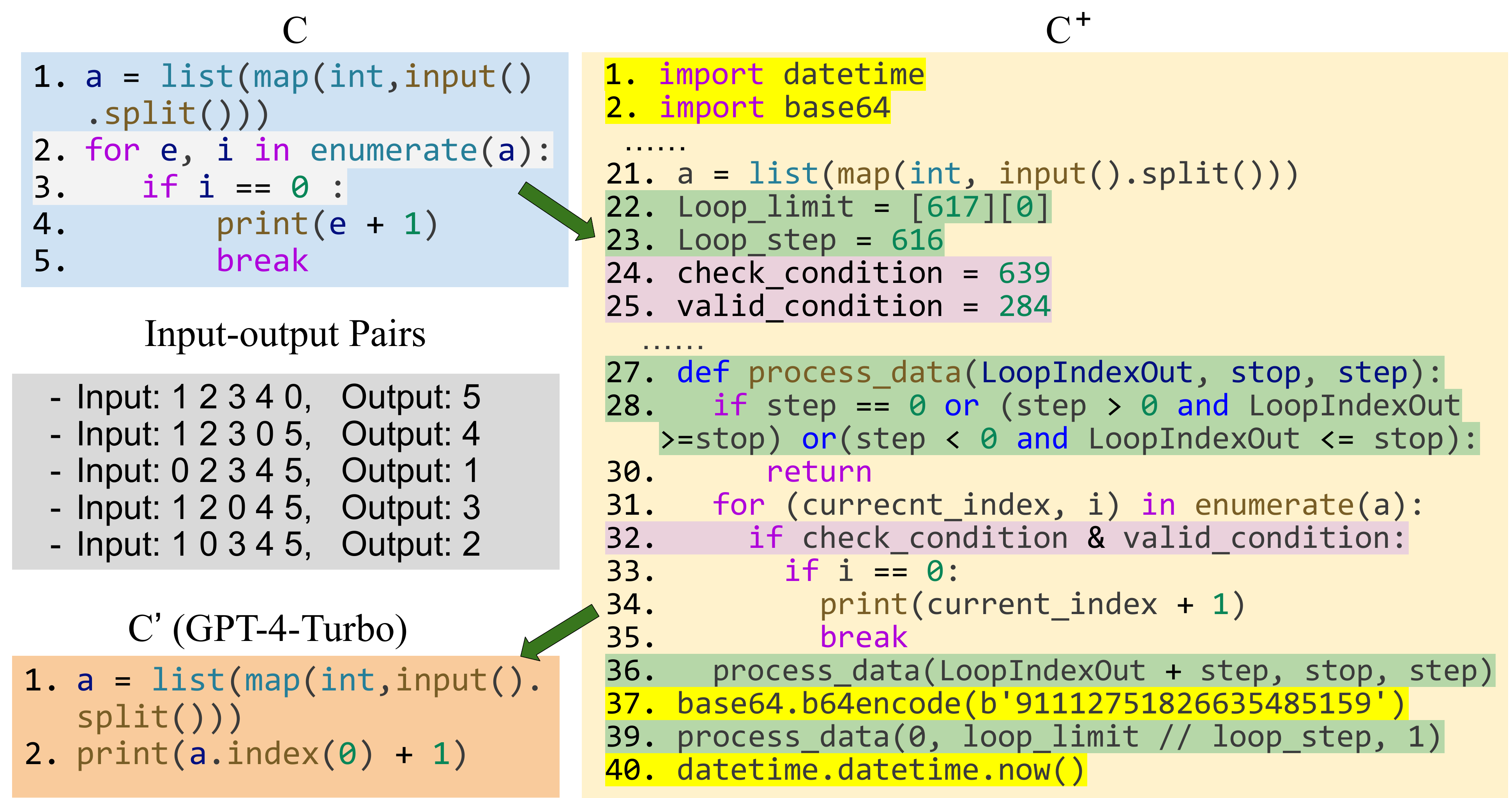}
    \vspace{-10pt}
    \caption{$C$, $C^+$, and $C^{\prime}$ for Avatar\_atcoder\_ABC170\_A. The input-output pairs pass on all three programs}
    \vspace{-10pt}
    \label{fig:cr-example}
\end{figure}

Each program will be \emph{reversely}\footnote{Given that the refactoring goal is to make code more readable, shorter, or optimized. We aimed to do the opposite, so we could not use existing refactoring tools and had to implement the reverse refactoring ourselves.} refactored multiple times, using a combination of applicable transformations, resulting in longer and complex semantically-preserving programs. To be fair to models, the \name's in-context example for this task teaches the model to refactor code containing the transformations. \textbf{Although this favors the models, and they can capture the refactoring patterns, applying the patterns and the combination of them is non-deterministic, making it challenging for the models, especially when the original programs are not simple}. Figure~\ref{fig:cr-example} shows an example of such transformations (yellow box) given the atcoder\_ABC170\_A program in Avatar (blue box). 
\input{Table/Table-Transformations}

Table~\ref{tab:CR} shows the results of this experiment. \textbf{Subject LLMs, on average, can refactor \revision{$66.61\%$} of these programs to semantically equivalent versions ($Pass@1(C^{\prime})$), achieving \revision{$63.58\%$} $R_{\CR}$ success (for \revision{$3.03\%$} of programs, LLMs generated longer code that will be automatically discarded per Equation~\ref{eq:SCSR})}. The frontier API-access LLMs outperform open-source LLMs, with average $R_{DSR}$ margins of \major{$31.00\%$ (\claude)}, \revision{$30.36\%$ (\omini)} from the best open-source model, \sem. 

Since DSR is an implicit reasoning task, we also studied the impact of dynamic factors on the results to make more informed conclusions: the \emph{size} and \emph{complexity} of the problem. To that end, we measured the size of $C^+$ in LoC for the former, and code similarity and Levenshtein distance between $C$ and $C^+$ for the latter. 
\revision{Figure \ref{fig:dsr}-a shows the distribution of LoC of $C$ and $C^+$, categorized by the success and failure of the LLM in the \CR task. On average, transformations increase the subject programs' size by $39.73$ lines, confirming the quality of transformations to challenge the models properly. \textbf{We observe that $C^+$ programs corresponding to successful \CR are slightly smaller compared to those corresponding to failures}. At the same time, the average size of $C^+$ remains below $50$ lines, far smaller than the context windows of studied LLMs (e.g., $10$K/$200$K for \codellama-Inst-34b/\omini). Therefore, despite the positive \emph{correlation}, the factor of size is unlikely to be the main \emph{cause} of success or failure.}


\revision{Figure~\ref{fig:dsr}-b demonstrates the normalized Levenshtein distance~\cite{yujian2007normalized} between $C$ and $C^+$ programs. The higher the distance, the more our pipeline has changed the original problem with the transformation. Since the Levenshtein distance can be biased toward longer programs, we also compare the Jaccard similarity
between $C$ and $C^+$ programs (Figure~\ref{fig:dsr}-c). For Jaccard similarity, a lower number indicates less similarity. These results suggest that LLMs tend to succeed on $C^+$s that are more similar to $C$s, confirming that problem difficulty is also an important factor impacting the \CR performance.}



\input{Table/Table-testing_similarity}

\begin{figure*}[t]
    \centering
    \includegraphics[width=0.95\linewidth]{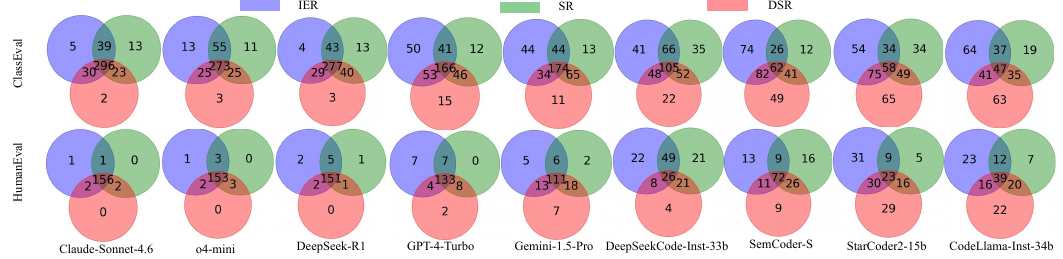}
    \vspace{-3pt}
    \caption{\major{Comparison of successful reasoning across \name's tasks}}
    \vspace{-10pt}
    \label{fig:comparison-reasoning}
\end{figure*}

\revision{Finally, we performed a Chi-square test~\cite{sloane1996introduction} with the following two null hypotheses: \emph{$H_0$. A higher Jaccard similarity between $C$ and $C^+$ is not associated with the success on \CR} and \emph{$H'_0$. A smaller LOC of $C^+$ is not associated with the success on \CR}.}
\revision{Table~\ref{tab:loc_similarity} demonstrates the P-values for different LLMs, confirming rejection of both null hypotheses (all P-values smaller than $0.05$). A smaller P-value generally indicates stronger evidence against the null hypothesis ($H_0$ and $H'_0$)~\cite{lehmann2005testing, barnard1949statistical,wasserstein2016asa}. As we can see, the P-values for $H_0$ are consistently lower than $H'_0$, indicating that the performance of LLMs on DSR is mainly impacted by the complexity of the problem, not its size, confirming a higher challenge for implicit code reasoning to identify and eliminate redundancies.}


From Table~\ref{tab:CR}, we observe that \textbf{there is no considerable correlation between the higher size or instruction-tuning and higher performance in \CR. We speculate this is because learning about code semantics has not been an explicit instruction-tuning objective, and given that the task is non-trivial, the ability to follow instructions better is not helpful}. Evidence for this claim is \sem, which is smaller than \deepseek-Inst-33b and \codellama-Inst-34b, but outperforms them with a considerable margin {$3.74\%$} (\deepseek-Inst-33b) and $9.33\%$ (\codellama-Inst-34b). \sem is fine-tuned on execution data; hence, it seems to better understand the code's general semantics and can more effectively identify and remove redundant statements.

\textbf{We observed cases where LLMs generated shorter programs than the original code (negative $LoC(C^{\prime})-LoC(C)$)}. Figure~\ref{fig:cr-example} presents such a case from \gptf simplifying transformed version of Avatar\_atcoder\_ABC170\_C. The transformation ($C^+$) surrounds the original for loop with a recursive function (Lines 27 and 28), adding a nested recursion in the code (Lines 36 and 39), which only runs for one time (\texttt{\small{LoopChecker12//LoopChecker22}} equals to 1). It also adds redundant API calls and their corresponding imports (Lines $1$, $2$, $37$, and $40$), and a conditional statement and corresponding boolean variables (Lines 24, 25, and 32). \gptf successfully identifies all these redundant statements and generates a semantically equivalent  $C^{\prime}$ that is three lines shorter than $C$: the model reduces the nested level of $C$ by replacing the \texttt{\small{For}} loop and the \texttt{\small{if}} statement with the \texttt{\small{.index()}} API of  \texttt{\small{list}}, which reflects its general knowledge of the programming language and code semantics of $C^\prime$.

%% file: Table/Table-CR.tex
\begin{table*}[]
\caption{\revision{Performance of subject LLMs in dynamic semantics reasoning measured by $R_{DSR}$ in Equation~\ref{eq:RCSR}. We highlight the top three best-performing models with {\color[HTML]{FE0000} red (\nth{1})}, {\color[HTML]{009901} green (\nth{2})}, and {\color[HTML]{3531FF} blue (\nth{3})}.}}
\vspace{-5pt}

\centering
\label{tab:CR}
\setlength{\tabcolsep}{1pt}
\resizebox{\textwidth}{!}{
\begin{tabular}{|c|c|ccccccccccccc|}
\hline
 &
   &
  \multicolumn{13}{c|}{\textbf{Subject LLMs}} \\ \cline{3-15} 
 &
   &
  \multicolumn{3}{c|}{\textbf{CodeLlama}} &
  \multicolumn{3}{c|}{\textbf{DeepSeek-Coder}} &
  \multicolumn{1}{c|}{} &
  \multicolumn{1}{c|}{} &
  \multicolumn{1}{c|}{} &
   \multicolumn{1}{c|}{} &
  \multicolumn{1}{c|}{} &
  \multicolumn{1}{c|}{} &
   \\ 
\multirow{-3}{*}{\textbf{Dataset}} &
  \multirow{-3}{*}{\textbf{Metrics}} &
  \multicolumn{1}{c|}{\textbf{(Inst-13b)}} &
  \multicolumn{1}{c|}{\textbf{(Base-13b)}} &
  \multicolumn{1}{c|}{\textbf{(Inst-34b)}} &
  \multicolumn{1}{c|}{\textbf{(Inst-6.7b)}} &
  \multicolumn{1}{c|}{\textbf{(Base-6.7b)}} &
  \multicolumn{1}{c|}{\textbf{(Inst-33b)}} &
  \multicolumn{1}{c|}{\multirow{-2}{*}{\textbf{\begin{tabular}[c]{@{}c@{}}\sem\\ (6.7b)\end{tabular}}}} &
  \multicolumn{1}{c|}{\multirow{-2}{*}{\textbf{\begin{tabular}[c]{@{}c@{}}StarCoder2\\ (15b)\end{tabular}}}} &
  \multicolumn{1}{c|}{\multirow{-2}{*}{\textbf{\begin{tabular}[c]{@{}c@{}}Gemini-1.5-\\ Pro\end{tabular}}}} &
  \multicolumn{1}{c|}{\multirow{-2}{*}{\textbf{\begin{tabular}[c]{@{}c@{}}GPT-4-\\ Turbo\end{tabular}}}} &
  \multicolumn{1}{c|}{\multirow{-2}{*}{\textbf{\begin{tabular}[c]{@{}c@{}}\major{Claude-Sonnet-}\\ \major{4.6}\end{tabular}}}} &
\multicolumn{1}{c|}{\multirow{-2}{*}{\textbf{\begin{tabular}[c]{@{}c@{}}\revision{DeepSeek-}\\ \revision{R1}\end{tabular}}}} &
  \multicolumn{1}{c|}{\multirow{-2}{*}{\textbf{\begin{tabular}[c]{@{}c@{}}\revision{o4-}\\ \revision{mini}\end{tabular}}}}
  
  \\ \hline
 &
  \textbf{Pass@1($C^{\prime}$)} &
  \multicolumn{1}{c|}{24.42\%} &
  \multicolumn{1}{c|}{24.42\%} &
  \multicolumn{1}{c|}{27.91\%} &
  \multicolumn{1}{c|}{20.93\%} &
  \multicolumn{1}{c|}{19.77\%} &
  \multicolumn{1}{c|}{24.42\%} &
  \multicolumn{1}{c|}{37.21\%} &
  \multicolumn{1}{c|}{{44.19\%}} &
  \multicolumn{1}{c|}{{60.47\%}} &
  \multicolumn{1}{c|}{{53.95\%}} &
  \multicolumn{1}{c|}{\color[HTML]{FE0000}94.74\%} &
  \multicolumn{1}{c|}{\color[HTML]{3531FF}88.16\%} &
  \multicolumn{1}{c|}{\color[HTML]{009901}90.79\%}
  \\ \cline{2-15} 
 \multirow{-2}{*}{\textbf{Avatar}} &
  \textbf{$R_{DSR}$} &
  \multicolumn{1}{c|}{$20.68\%$} &
  \multicolumn{1}{c|}{$19.74\%$} &
  \multicolumn{1}{c|}{$22.37\%$} &
  \multicolumn{1}{c|}{$17.54\%$} &
  \multicolumn{1}{c|}{$18.11\%$} &
  \multicolumn{1}{c|}{$18.22\%$} &
  \multicolumn{1}{c|}{$30.84\%$} &
  \multicolumn{1}{c|}{{$34.51\%$}} &
  \multicolumn{1}{c|}{{$53.74\%$}} &
  \multicolumn{1}{c|}{{$48.92\%$}} &
  \multicolumn{1}{c|}{\color[HTML]{FE0000} 88.23\%} &
    \multicolumn{1}{c|}{\color[HTML]{3531FF}76.88\%} &
  \multicolumn{1}{c|}{\color[HTML]{009901}85.22\%}
  \\ \cline{2-15} 
 \hline
 &
  \textbf{Pass@1($C^{\prime}$)} &
  \multicolumn{1}{c|}{45.85\%} &
  \multicolumn{1}{c|}{40.50\%} &
  \multicolumn{1}{c|}{47.50\%} &
  \multicolumn{1}{c|}{60.75\%} &
  \multicolumn{1}{c|}{{56.00\%}} &
  \multicolumn{1}{c|}{{68.50\%}} &
  \multicolumn{1}{c|}{{60.50\%}} &
  \multicolumn{1}{c|}{59.25\%} &
  \multicolumn{1}{c|}{{80.45\%}} &
  \multicolumn{1}{c|}{{79.32\%}} &
  \multicolumn{1}{c|}{\color[HTML]{FE0000}90.93\%} &
\multicolumn{1}{c|}{\color[HTML]{009901} 90.42\%} &
  \multicolumn{1}{c|}{\color[HTML]{3531FF} 84.46\%}
  \\ \cline{2-15} 
 \multirow{-2}{*}{\textbf{ClassEval}} &
  \textbf{$R_{DSR}$} &
  \multicolumn{1}{c|}{$42.21\%$} &
  \multicolumn{1}{c|}{$38.03\%$} &
  \multicolumn{1}{c|}{$46.04\%$} &
  \multicolumn{1}{c|}{$57.90\%$} &
  \multicolumn{1}{c|}{$54.62\%$} &
  \multicolumn{1}{c|}{{$65.47\%$}} &
  \multicolumn{1}{c|}{{$56.85\%$}} &
  \multicolumn{1}{c|}{$56.86\%$} &
  \multicolumn{1}{c|}{{$77.51\%$}} &
   \multicolumn{1}{c|}{{$76.98\%$}}&
   \multicolumn{1}{c|}{\color[HTML]{FE0000} 84.71\%} &
  \multicolumn{1}{c|}{ \color[HTML]{009901} 83.45\%} &
  \multicolumn{1}{c|}{ \color[HTML]{3531FF} 81.52\%}
  \\ \cline{2-15} 
  \hline
 &
  \textbf{Pass@1($C^{\prime}$)} &
  \multicolumn{1}{c|}{72.25\%} &
  \multicolumn{1}{c|}{70.25\%} &
  \multicolumn{1}{c|}{75.13\%} &
  \multicolumn{1}{c|}{76.77\%} &
  \multicolumn{1}{c|}{65.38\%} &
  \multicolumn{1}{c|}{{83.29\%}} &
  \multicolumn{1}{c|}{78.13\%} &
  \multicolumn{1}{c|}{79.00\%} &
  \multicolumn{1}{c|}{{80.17\%}} &
  \multicolumn{1}{c|}{{86.13\%}} &
  \multicolumn{1}{c|}{\color[HTML]{009901} 97.38\%} &
  \multicolumn{1}{c|}{\color[HTML]{3531FF} 97.00\%} &
  \multicolumn{1}{c|}{ \color[HTML]{FE0000} 98.75\%}
  \\ \cline{2-15} 
 \multirow{-2}{*}{\textbf{CRUXEval}} &
  \textbf{$R_{DSR}$} &
  \multicolumn{1}{c|}{$72.11\%$} &
  \multicolumn{1}{c|}{$70.18\%$} &
  \multicolumn{1}{c|}{$74.38\%$} &
  \multicolumn{1}{c|}{$77.02\%$} &
  \multicolumn{1}{c|}{$65.28\%$} &
  \multicolumn{1}{c|}{{$79.35\%$}} &
  \multicolumn{1}{c|}{{$78.03\%$}} &
  \multicolumn{1}{c|}{$77.55\%$} &
  \multicolumn{1}{c|}{$78.00\%$} &
  \multicolumn{1}{c|}{{$85.91\%$}} &
  \multicolumn{1}{c|}{\color[HTML]{3531FF} 92.50\%} &
    \multicolumn{1}{c|}{\color[HTML]{009901}93.59\%} &
  \multicolumn{1}{c|}{\color[HTML]{FE0000} 98.09\%}
  \\ \cline{2-15} 
  \hline
 &
  \textbf{Pass@1($C^{\prime}$)} &
  \multicolumn{1}{c|}{51.83\%} &
  \multicolumn{1}{c|}{37.20\%} &
  \multicolumn{1}{c|}{60.98\%} &
  \multicolumn{1}{c|}{60.37\%} &
  \multicolumn{1}{c|}{40.24\%} &
  \multicolumn{1}{c|}{64.02\%} &
  \multicolumn{1}{c|}{{75.00\%}} &
  \multicolumn{1}{c|}{64.63\%} &
  \multicolumn{1}{c|}{{91.98\%}} &
  \multicolumn{1}{c|}{{90.74\%}} &
  \multicolumn{1}{c|}{\color[HTML]{FE0000} 98.77\%} &
\multicolumn{1}{c|}{\color[HTML]{3531FF} 95.06\%} &
  \multicolumn{1}{c|}{\color[HTML]{009901} 97.53\%}
  \\ \cline{2-15} 
 \multirow{-2}{*}{\textbf{HumanEval}} &
  \textbf{$R_{DSR}$} &
  \multicolumn{1}{c|}{$51.57\%$} &
  \multicolumn{1}{c|}{$36.65\%$} &
  \multicolumn{1}{c|}{$59.15\%$} &
  \multicolumn{1}{c|}{$59.71\%$} &
  \multicolumn{1}{c|}{$35.91\%$} &
  \multicolumn{1}{c|}{$61.29\%$} &
  \multicolumn{1}{c|}{{$73.55\%$}} &
  \multicolumn{1}{c|}{$64.25\%$} &
  \multicolumn{1}{c|}{{$89.87\%$}} &
  \multicolumn{1}{c|}{{$89.04\%$}} &
  \multicolumn{1}{c|}{\color[HTML]{FE0000} 97.83\%} &
 \multicolumn{1}{c|}{\color[HTML]{3531FF} 92.42\%} &
  \multicolumn{1}{c|}{\color[HTML]{009901} 95.87\%} 
  
  \\ \cline{2-15} 
  \hline

  \rowcolor{lightgray} \multicolumn{2}{|c|}{\textbf{$\rho_{CC}$}} &
  \multicolumn{1}{c|}{-0.51} &
  \multicolumn{1}{c|}{-0.39} &
  \multicolumn{1}{c|}{-0.46} &
  \multicolumn{1}{c|}{-0.62} &
  \multicolumn{1}{c|}{-0.57} &
  \multicolumn{1}{c|}{-0.61} &
  \multicolumn{1}{c|}{-0.88} &
  \multicolumn{1}{c|}{-0.83} &
  \multicolumn{1}{c|}{-0.81} &
  \multicolumn{1}{c|}{-0.87} &
  \multicolumn{1}{c|}{-0.56} &
  \multicolumn{1}{c|}{-0.74} &
  \multicolumn{1}{c|}{-0.68}
  \\ \hline

\rowcolor{lightgray} \multicolumn{2}{|c|}{\textbf{$\rho_{LoC}$}} &
  \multicolumn{1}{c|}{-0.43} &
  \multicolumn{1}{c|}{-0.62} &
  \multicolumn{1}{c|}{-0.70} &
  \multicolumn{1}{c|}{-0.61} &
  \multicolumn{1}{c|}{-0.53} &
  \multicolumn{1}{c|}{-0.63} &
  \multicolumn{1}{c|}{-0.89} &
  \multicolumn{1}{c|}{-0.66} &
  \multicolumn{1}{c|}{-0.53} &
  \multicolumn{1}{c|}{-0.67} &
  \multicolumn{1}{c|}{-0.77} &
   \multicolumn{1}{c|}{-0.70} &
  \multicolumn{1}{c|}{-0.57}
  \\ \hline

\rowcolor{lightgray}   \multicolumn{2}{|c|}{\textbf{$\rho_{DEP}$}} &
  \multicolumn{1}{c|}{-0.40} &
  \multicolumn{1}{c|}{-0.36} &
  \multicolumn{1}{c|}{-0.58} &
  \multicolumn{1}{c|}{-0.49} &
  \multicolumn{1}{c|}{-0.42} &
  \multicolumn{1}{c|}{-0.68} &
  \multicolumn{1}{c|}{-0.69} &
  \multicolumn{1}{c|}{-0.71} &
  \multicolumn{1}{c|}{-0.61} &
  \multicolumn{1}{c|}{-0.82} &
  \multicolumn{1}{c|}{-0.69} &
   \multicolumn{1}{c|}{-0.77} &
  \multicolumn{1}{c|}{-0.81}
  \\ \hline

\rowcolor{lightgray}   \multicolumn{2}{|c|}{\textbf{$\rho_{NC}$}} &
  \multicolumn{1}{c|}{-0.28} &
  \multicolumn{1}{c|}{-0.21} &
  \multicolumn{1}{c|}{-0.39} &
  \multicolumn{1}{c|}{-0.37} &
  \multicolumn{1}{c|}{-0.44} &
  \multicolumn{1}{c|}{-0.58} &
  \multicolumn{1}{c|}{-0.41} &
  \multicolumn{1}{c|}{-0.69} &
  \multicolumn{1}{c|}{-0.29} &
  \multicolumn{1}{c|}{-0.67} &
  \multicolumn{1}{c|}{-0.58} &
   \multicolumn{1}{c|}{-0.64} &
  \multicolumn{1}{c|}{-0.63}  
  
  \\ \hline

\rowcolor{lightgray}   \multicolumn{2}{|c|}{\textbf{$\rho_{LL}$}} &
  \multicolumn{1}{c|}{-0.50} &
  \multicolumn{1}{c|}{-0.32} &
  \multicolumn{1}{c|}{-0.46} &
  \multicolumn{1}{c|}{-0.52} &
  \multicolumn{1}{c|}{-0.56} &
  \multicolumn{1}{c|}{-0.12} &
  \multicolumn{1}{c|}{-0.78} &
  \multicolumn{1}{c|}{-0.27} &
  \multicolumn{1}{c|}{-0.16} &
  \multicolumn{1}{c|}{-0.21} &
  \multicolumn{1}{c|}{-0.40} &
  \multicolumn{1}{c|}{-0.37} &
  \multicolumn{1}{c|}{-0.48}    
  \\ \hline

\end{tabular}
}
\vspace{-5pt}
\end{table*}

%% file: Table/Table-Transformations.tex
\begin{table}[]
\caption{Transformation Rules.}
\vspace{-5pt}
\label{tab:transformation}
\centering
\scalebox{0.9}{
\begin{tabular}{|c|c|}
\hline
\textbf{Type} & \textbf{Transformation}                              \\ \hline
\multirow{7}{*}{\textbf{\begin{tabular}[c]{@{}c@{}}Code\\ Structure\end{tabular}}} & Add another nested for to existing for loop \\ \cline{2-2} 
              & Add another nested if to existing if statement       \\ \cline{2-2} 
              & Add another nested while to existing while loop      \\ \cline{2-2} 
              & Introduce a thread to existing function call         \\ \cline{2-2} 
              & Add a try-except handler inside existing functions   \\ \cline{2-2} 
              & Replace applicable built-in calculations with Numpy  \\ \cline{2-2} 
              & Transform augment assignment                         \\ \cline{2-2} 
              & Transform existing for loop into recursive functions \\ \hline
\multirow{8}{*}{\textbf{\begin{tabular}[c]{@{}c@{}}API\\ Calls\end{tabular}}}       & Introduce API calls from base64 library     \\ \cline{2-2} 
              & Introduce API calls from cryptography library        \\ \cline{2-2} 
              & Introduce API calls from datetime library            \\ \cline{2-2} 
              & Introduce API calls from dateutil library            \\ \cline{2-2} 
              & Introduce http connections                           \\ \cline{2-2} 
              & Introduce API calls from scipy library               \\ \cline{2-2} 
              & Introduce API calls from sklearn library             \\ \cline{2-2} 
              & Introduce API calls from time library                \\ \hline
\multirow{2}{*}{\textbf{\begin{tabular}[c]{@{}c@{}}Inter/Intra-Procedural \\ Dependencies\end{tabular}}}
              & Transform existing statements into new functions       \\ \cline{2-2} 
              & Introduce a decorator     \\ \hline
\multirow{2}{*}{\textbf{Renaming}}                                                  & Rename existing variables                   \\ \cline{2-2} 
              & Rename existing functions                            \\ \hline
\end{tabular}
}
\vspace{-10pt}
\end{table}

%% file: Table/Table-testing_similarity.tex
\begin{table*}[t]
\caption{\revision{Chi-square test results for $H_0$ and $H'_0$.}}
\label{tab:loc_similarity}
\scalebox{0.80}{
\begin{tabular}{|c|c|c|c|c|c|c|c|c|c|}
\hline
\textbf{} &
  \textbf{\major{Claude-Sonnet-4.6}} &
  \textbf{o4-mini} &
  \textbf{DeepSeek-R1} &
  \textbf{GPT-4-Turbo} &
  \textbf{Gemini-1.5-Pro} &
  \textbf{DeepSeek-Coder-Inst-33b} &
  \textbf{SemCoder-S} &
  \textbf{StarCoder2-15b} &
  \textbf{CodeLlama-Inst-34b} \\ \hline
\textbf{$H_0$} &
  6.02E-7 &
  2.65E-8 &
  3.31E-9 &
  4.47E-9 &
  6.87E-9 &
  1.21E-10 &
  4.23E-8 &
  4.54E-9 &
  1.14E-12 \\ \hline
\textbf{$H_0^\prime$} &
  4.53E-4&
  2.13E-3 &
  2.40E-5 &
  1.46E-7 &
  6.26E-5 &
  4.04E-6 &
  5.76E-5 &
  1.20E-4 &
  3.38E-7 \\ \hline
\end{tabular}
}
\end{table*}

%% file: sections/4-4-Factors.tex
\vspace{-8pt}
\subsection{RQ4: Analysis of Reasoning Failures}
\label{subsec:eval-factors}

We have developed \exerscope~\cite{liu2025tool,exerscope} tool under the \name framework that can be plugged into any code reasoning framework and automatically assesses the impact of different (1) program constructs, (2) program complexities, (3) dynamic programming properties such as recursion length, and (4) variable types on code reasoning abilities of LLMs. \textbf{Analyzing the results of \name's three reasoning tasks with \exerscope shows that recursive and nested program constructs, longer loop iterations, and non-primitive types \emph{negatively} impact the reasoning ability of LLMs}.

\exerscope results also confirm the generalizability of our speculations in previous RQs (\S \ref{subsec:eval-IER}--\S \ref{subsec:eval-DSR}), i.e., the negative impact of program complexity on code reasoning performance, by measuring the Spearman's Rank Order Correlation (ROC)~\cite{spearman1961proof} between five different complexity metrics introduced in \S \ref{sec:setup}, and the $R_{\IER}$, $R_{\SR}$, and $R_{\CR}$ values. The calculated $\rho$ values are reported under the last five rows of Tables~\ref{tab:IER}--\ref{tab:CR} (highlighted in gray)\footnote{For \CR, \exerscope collects the CC (cyclomatic complexity) of the transformed programs ($C^+$) since they are directly exposed to LLMs.}. 
Except for a few cases with a slight negative correlation (e.g., \deepseek-Inst-33b $\langle R_{\CR}$, LL$\rangle$ or \starcoder $\langle S_{SR}$, LoC$\rangle$), \textbf{there is always a moderate to a strong negative correlation between complexity metrics and code reasoning performances of models}, confirming the struggle of LLMs to deal with complex code. For \IER and \SR, the impact of \emph{intra-class dependency} is strongest. For \CR, a higher \emph{cyclomatic complexity} makes the task more challenging. This is mainly because \IER and \SR require \textbf{simulating one execution path by LLMs}, and a longer path challenges their memorization and attention more. On the contrary, for \CR, LLMs should \textbf{simulate multiple execution paths} to understand the whole code semantics; thereby, more execution paths challenge the models more.

%% file: sections/4-5-Different-Reasoning-Tasks.tex
\subsection{RQ5: Necessity for Different Code Reasoning Tasks}
\label{subsec:eval-different-reasoning}

Prior techniques, such as \crux and \reval, focus on explicit execution reasoning, while \name proposes two new tasks and metrics that entail execution awareness but require different aspects of code semantics understanding. To show the necessity of including implicit code reasoning tasks, we investigate whether explicit code execution reasoning (\IER) subsumes the other two tasks (\SR and \CR). 

We examined the overlap between the programs that individual LLMs correctly reasoned about under different code reasoning tasks. Since \SR was evaluated only on \heval and \ceval, this experiment considers the program in these two benchmarks for a fair comparison. Figure~\ref{fig:comparison-reasoning} shows the results of this experiment (for \deepseek and \codellama, we selected the best-performing model in the family).  
We can see that, \textbf{on both explicit and implicit reasoning tasks, \major{\claude}, \revision{\omini}, and \revision{\dsr} consistently yield higher correct predictions for \major{$80.14\%$},  \revision{$75.53\%$} and \revision{$75.89\%$} of the studied programs, respectively. For other models, the overlap becomes less prevalent}. For example, \deepseek-Inst-33b achieves correct predictions on $23.23\%$ of the programs across all three reasoning tasks, and the percentage decreases to $15.25\%$ for \codellama-Inst-34b.
This study also shows that \textbf{while there is an overlap between the successful cases of the three tasks, the number of programs exclusive to each reasoning task is considerable}. 

\begin{figure}
    \centering
    \includegraphics[width=0.73\linewidth]{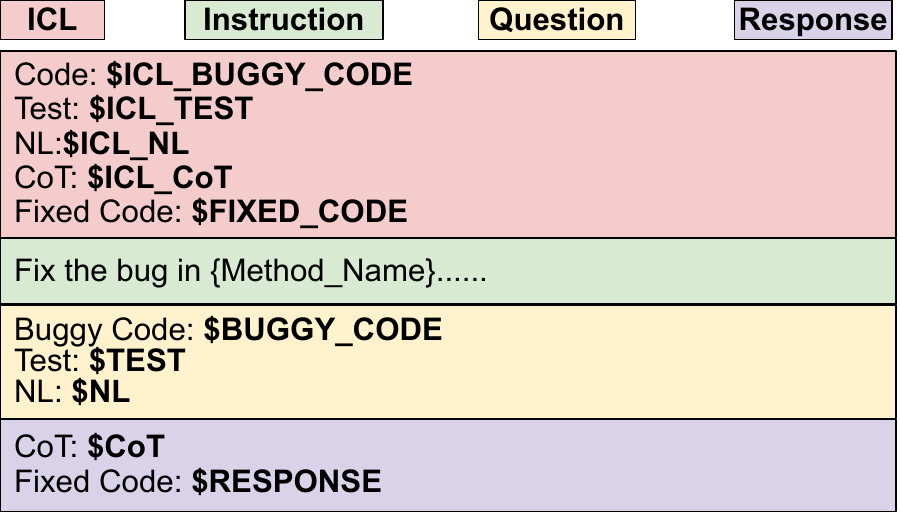}
    \caption{Prompt template used for Bug Repair (BR)}
    \vspace{-20pt}
    \label{fig:prompt-repair}
\end{figure}

\begin{figure*}[t]
    \centering
    \includegraphics[width=0.99\linewidth]{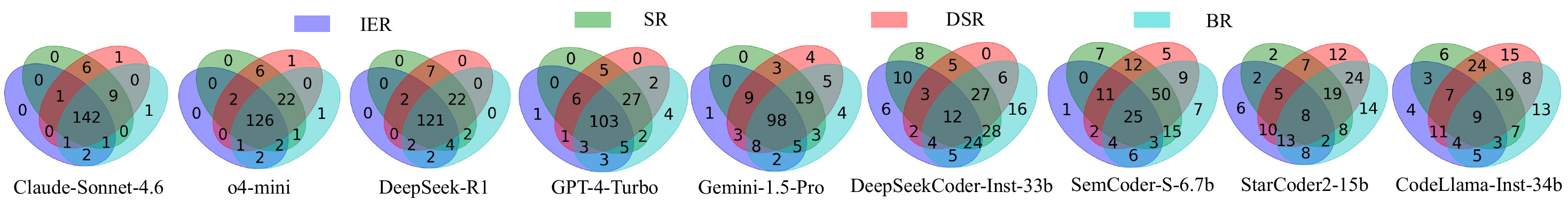}
    \caption{ \major{Correct predictions of LLMs on IER, SR, DSR, and BR tasks}}
    \vspace{-5pt}
    \label{fig:rq5-venn}
\end{figure*}

These results confirm \textbf{the necessity of evaluating LLMs with different reasoning tasks rather than focusing only on execution reasoning}. These results also show that \textbf{execution awareness does not necessarily improve the implicit code reasoning abilities of the model, serving as a guideline to model developers to incorporate implicit reasoning to account for the next generation of code LLMs}. We believe \name is just the beginning, and more tasks can be designed on top of it to assess other aspects of code reasoning. 

%% file: sections/4-6-APR-Comparison.tex
\vspace{-15pt}
\subsection{RQ6: Code Reasoning and Bug Repair}
\label{subsec:eval-APRandReasoning}

Over the past years, many programming tasks have been proposed to evaluate the programming abilities of LLMs. Intuitively, LLMs should understand the programming languages and incorporate this knowledge and the code examples they have seen during training to perform the programming tasks. Therefore, one can claim that LLMs are already being evaluated for code reasoning. To understand whether this intuition holds or if there is a need for code reasoning tasks, we compare LLMs' performance on Bug Repair (BR) with their performance on the three code reasoning tasks of \name. Bug Repair is a programming task that requires a deep understanding of code semantics: it should understand the semantics of buggy code with respect to the specifications and tests, and generate a patch accordingly. 
Thus, we define the following expectations: (1) if BR already evaluates the code reasoning, the model should pass the code reasoning tasks for successful bug repair cases; (2) if the model cannot repair a buggy code, it is likely because of the reasoning failure. 

To further investigate whether LLMs meet the two expectations above, we used HumanEvalPack~\cite{muennighoff2023octopack}, a dataset of bugs generated by humans and injected into HumanEval programs.
We identified error-revealing tests from this dataset, i.e., those that pass on the correct code but fail on the buggy code. Then, we repeated our experiments in the first three research questions, asking LLMs to perform \IER, \SR, and \CR on the buggy code considering the error-revealing tests. We also asked subject LLMs to repair the bugs along with test information using the prompt presented in Figure~\ref{fig:prompt-repair}.

\input{Table/Table-APR}

Table~\ref{tab:agreement_prog_tasks} illustrates the performance of the LLMs in repairing the bugs and code reasoning tasks (for \deepseek and \codellama, we selected the best-performing model in the family). The Venn diagrams in Figure \ref{fig:rq5-venn} also visualize the successful cases under different tasks, emphasizing the unique cases and overlaps.
We can see that \textbf{\claude, \omini, \dsr are capable of making correct predictions on all four tasks on \major{$90.44\%$}, $80.25\%$,  and $77.07\%$ of \heval problems, respectively.} However, there is less overlap in other LLMs, especially those open-access non-reasoning models: for example, only $5.73\%$ programs fall into the overlap of four tasks for \codellama-Inst-34b. These results show that \textbf{prior to reasoning models, which are specifically trained to bring the reasoning process into problem solving, there was no meaningful association between the abilities of the models in code reasoning tasks and bug repair.} However, reasoning-enabled models, although not specifically trained for \name reasoning tasks, not only excel in them, but also incorporate explicit and implicit code reasoning into problem solving for other programming tasks. 




\begin{figure}
    \centering
    \vspace{-10pt}\includegraphics[width=0.95\linewidth]{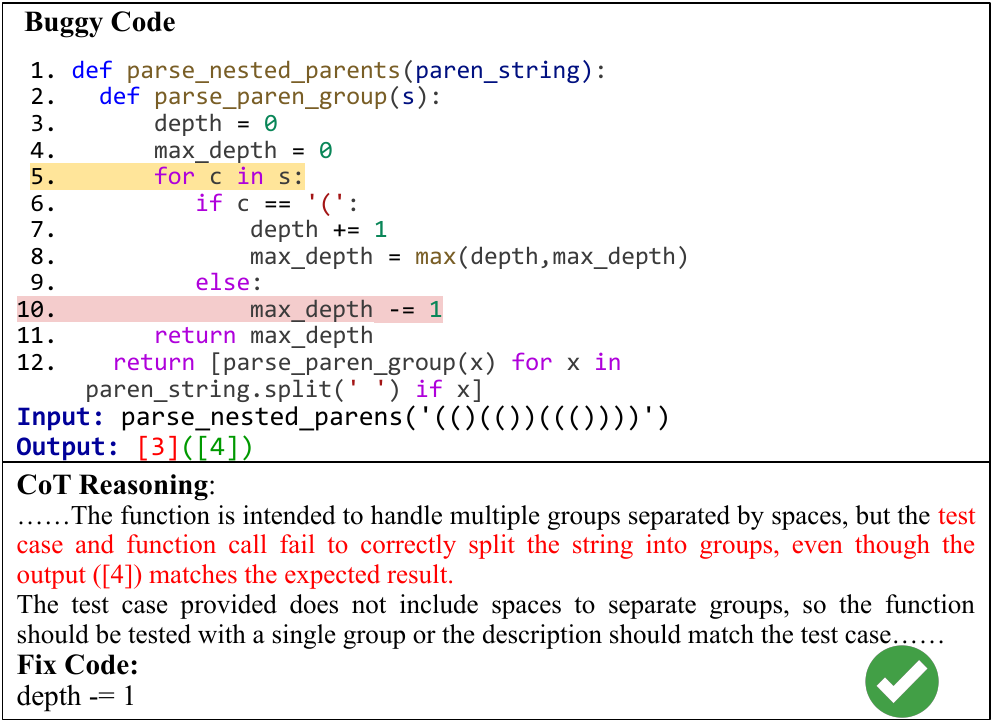}
    \caption{\major{\gptf takes a reasoning shortcut on HumanEval/6}}
    \label{fig:shortcut}
    \vspace{-10pt}
\end{figure}

\begin{figure*}[t]
    \centering
    \includegraphics[width=0.99\linewidth]{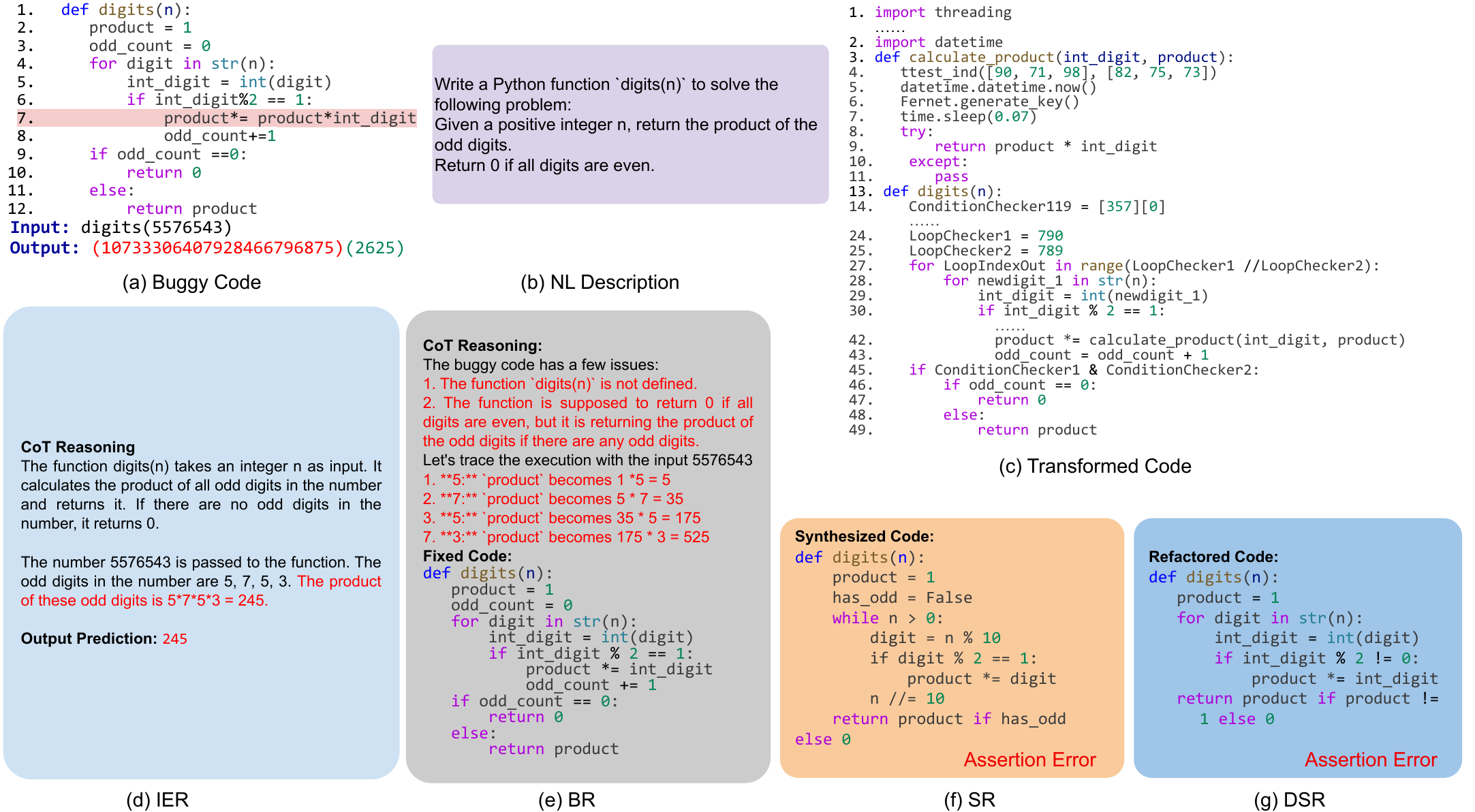}
    \caption{An example of incorrect \IER (d), \SR (f), and \CR (g) by \gemini for HumanEval/131, and correct BR (e)}
    \vspace{-12pt}
    \label{fig:rq5-correct-apr}
\end{figure*}

\revision{
These results also raise a question: \emph{if an LLM, reasoning or non-reasoning, succeeds in bug repair without success in reasoning, what could be the reasons?} To that end, we performed a quantitative analysis of successful Bug Repair (BR) predictions across all evaluated LLMs categorized in two groups: (a) problems where the LLM fails all three code-reasoning tasks (“All Incorrect” in Table~\ref{tab:correct_br}), and (b) problems where the LLM fails one or two of them (“At least one Incorrect” in Table~\ref{tab:correct_br}). For category (b), we randomly sample instances for each LLM at a $95\%$ confidence level for manual inspection and report the number of sampled cases in Table~\ref{tab:correct_br} (column “\#Sampling”). We observed that, in $45.00\%$ and $39.58\%$ of the manually inspected cases from categories (a) and (b), the success in BR was due to natural language shortcuts and lucky hallucinations.}

\major{In manual analysis, a reasoning shortcuts are where LLM bypasses \emph{any} detailed execution-related analysis of the test input and directly asserts the output (whether correct or incorrect). This behavior is distinct from a valid logic leap, where the model may omit one or two intermediate steps while still presenting coherent, execution-based reasoning. Figure~\ref{fig:shortcut} illustrates such an example: \gptf asserts the result for the test without reasoning through the program’s execution in its CoT, and its asserted output ([4]) is inconsistent with the actual execution. However, it still produces a correct bug fix despite relying on an incorrect and incomplete reasoning process. We speculate that it is because the model infers the intended behavior from the method signature (\texttt{parse\_nested\_group}) and uses this to implement the fix.}

\input{Table/Table-suspicious}

Figure~\ref{fig:rq5-correct-apr} presents a case, where \gemini successfully repairs the bug but fails on all three explicit/implicit code reasoning tasks. The bug is in line 7 in Figure~\ref{fig:rq5-correct-apr}-a, where it incorrectly computes the \emph{product of the odd digits} specified in Figure~\ref{fig:rq5-correct-apr}-b. As a result, given the input \texttt{\small{digits(5576543)}}, the code returns a very large number instead of $2625$.
From Figure~\ref{fig:rq5-correct-apr}-d, we can see that \gemini \major{does not} understand the buggy line and fails to follow the execution of the buggy code: it fails to predict which \texttt{int\_digit} satisfies the if condition in line $6$ and misunderstands the statement in line $7$. Similarly, it fails to generate the correct code in Figure~\ref{fig:rq5-correct-apr}-f and Figure~\ref{fig:rq5-correct-apr}-g. From Figure~\ref{fig:rq5-correct-apr}-e, we can see that \gemini is capable of repairing the bug; however, the correct fix is based on the incorrect reasoning: (1) \gemini incorrectly identifies the bug location, (2) it also incorrectly simulates the execution process of the test case. It fails to track the state of \texttt{\small{int\_digit}} in line 5 and line 7, which should be $[5,5,7,5,3]$ instead of $[5,7,5,3]$.

%% file: Table/Table-APR.tex
\begin{table}[]
\caption{\major{Evaluating LLMs' performance on  Bug Repair (BR) task and \name's reasoning tasks.}}
\vspace{-5pt}
\centering
\label{tab:agreement_prog_tasks}
\scalebox{0.95}{
\begin{tabular}{|c|c|c|c|c|}
\hline
\multicolumn{1}{|l|}{\textbf{}}     & \textbf{IER} & \textbf{SR} & \textbf{DSR} & \textbf{BR} \\ \hline
\textbf{CodeLlama-Inst-34b}     & 29.30\%      & 47.78\%     & 57.93\%      & 43.31\%     \\ \hline
\textbf{DeepSeekCoder-Inst-33b} & 42.04\%      & 71.34\%     & 60.78\%      & 77.71\%     \\ \hline
\textbf{SemCoder-S-6.7b}            & 33.12\%      & 75.16\%     & 75.00\%      & 75.80\%     \\ \hline
\textbf{StarCoder2-15b}             & 34.39\%      & 49.39\%     & 62.20\%      & 61.15\%     \\ \hline
\textbf{Gemini-1.5-Pro}             & 75.80\%      & 85.99\%     & 91.46\%      & 91.72\%     \\ \hline
\textbf{GPT-4-Turbo}                & 77.71\%      & 91.08\%     & 90.74\%      & 94.90\%      \\ \hline
\textbf{\revision{DeepSeek-R1}}                & 83.44\%      & 96.82\%     & 95.06\%      & 96.18\%   \\ \hline
\textbf{\revision{o4-mini}}                & 84.71\%      & 97.45\%     & 92.42\%      & 98.09\% \\ \hline

\textbf{\major{Claude-Sonnet-4.6}}                & 93.63\%      & 98.09\%     & 96.82\%      & 99.36\%
\\ \hline
\end{tabular}
}
\vspace{-10pt}
\end{table}

%% file: Table/Table-suspicious.tex
\begin{table}[]
\caption{\revision{Number of identified suspicious success (\#Sus) in BR.}}
\label{tab:correct_br}
\scalebox{0.95}{
\begin{tabular}{|c|cc|cc|}
\hline
\multirow{2}{*}{} & \multicolumn{2}{c|}{\textbf{All Incorrect}}                        & \multicolumn{2}{c|}{\textbf{At least one Incorrect}}                \\ \cline{2-5} 
                  & \multicolumn{1}{c|}{\textbf{\#Sampling}} & \textbf{\#Sus} & \multicolumn{1}{c|}{\textbf{\#Sampling}} & \textbf{\#Sus} \\ \hline
\textbf{CodeLlama-Inst-34b}     & 13 &    7      & 32  &     15      \\
\textbf{DeepSeekCoder-Inst-33b} & 16 &    9       & 47  &   22        \\
\textbf{SemCoder-S-6.7b}        & 7  &   3        & 46  &   18        \\
\textbf{StarCoder2-15b}         & 14 &   5        & 43  &   19        \\
\textbf{Gemini-1.5-Pro}         & 4  &  1        & 30  &    8       \\
\textbf{GPT-4-Turbo}            & 4  &  2        & 30  &   10        \\
\textbf{DeepSeek-R1}            & 0  &  0         & 25  & 14          \\
\textbf{o4-mini}                & 1  &   0        & 22  & 6          \\ 
\textbf{\major{Claude-Sonnet-4.6}}      & 1  &   0        & 13  & 2          \\ 
\hline
\textbf{Total}                  & 60 & 27 & 288 & 114 \\ \hline
\end{tabular}
}
\vspace{-20pt}
\end{table}

%% file: sections/4-7-Agentic-System.tex
\subsection{\major{RQ7: Code Reasoning in Programming Agents}}
\label{subsec:agentic}

\major{A prominent research trend in software engineering is the adoption of agentic systems equipped with diverse tools to address increasingly challenging tasks on real-world codebases.}
\major{This trend raises a critical question: \emph{Given the capability of tool usage, is code reasoning required for them to perform?}
To answer this question, we obtained $2,000$ trajectories from Liu et al.~\cite{liu2025process}, generated by running SWE-agent~\cite{yang2024swe} on $500$ instances of SWE-Bench-Verified~\cite{jimenez2023swe} using four models: \dsr, Claude-Sonnet-4, Devstral~\cite{mistral2025devstral2507}, and DeepSeek V3~\cite{liu2024deepseek}. The study used default system prompts and settings of SWE-agent, without instructing the agent to reason about code execution during localization, patching, or validation.}

\major{
Given the substantial human effort required to manually inspect natural-language reasoning traces, and the limited reliability of LLM-as-a-judge methods~\cite{wang2025trustjudge, bavaresco2025llms, stureborg2024large}, we adopt a lightweight yet effective keyword-based heuristic to identify trajectories that appear to perform execution-aware reasoning \emph{before} taking an action (including external tool use). Specifically, we searched the agent's \emph{thought} field at each trajectory step for phrases indicative of execution reasoning, such as \emph{evaluates to}, \emph{the value of}, \emph{let me trace}, and \emph{when we call}.
Across the studied trajectories, we identified $419$ in which the agent potentially incorporates code reasoning into the decision-making process.
}

\begin{figure}
    \centering
    \includegraphics[width=0.5\textwidth]{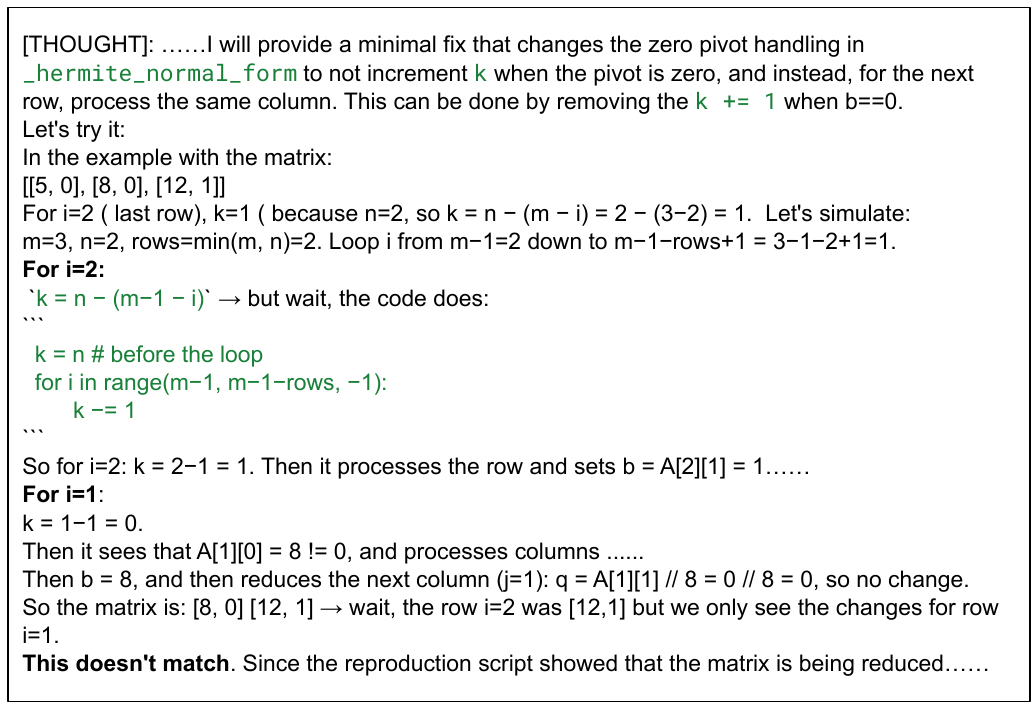}
    \caption{\major{The reasoning process of SWE-agent (DeepSeek-R1) at trajectory step 9 for sympy-23413}}
    \label{fig:agent_thought-middle-correction}
    \vspace{-15pt}
\end{figure}

\major{Figure~\ref{fig:agent_thought} presents one such example, in which the agent explicitly simulates the program’s execution within its reasoning process after an edit from the previous step fails the test case, and before applying subsequent edits.
Figure~\ref{fig:agent_thought-middle-correction} provides another example, where the agent simulates the execution of a proposed fix: \dsr traces loop iterations over a matrix using concrete values (\texttt{\small{m=3,n=2}}), accesses matrix elements by index, and tracks the matrix state across iterations. Notably, through this internal simulation, the agent determines that the proposed fix would yield behavior inconsistent with the expected output, and then pivots to an alternative workaround.}

\major{These cases suggest that implicit or explicit code reasoning has been embedded in the planning and decision-making process of LLM agents, even when external execution tools are available to run the code.} 

%% file: sections/4-8-REval-Comparison.tex
\vspace{-8pt}
\subsection{RQ8: Comparison with Alternative Approaches}
\label{subsec:comparison}

We evaluated \name's \IER task with the most recent related work, \reval, which evaluates LLMs using four runtime behavior prediction tasks: for given inputs and a statement in the program, \reval prompts LLMs to predict (1) if the statement is covered during execution, (2) variable values after the execution of it, (3) the next statement to be executed after it, and (4) the final output. Specifically, we compared the output prediction results for the common programs and studied LLMs in the two techniques. \reval is evaluated using a subset of the programs in \heval and \ceval. We identified those programs and extracted the outcome of LLMs for output prediction from \reval's artifacts to compare the results with those of \name.

Table~\ref{tab:reval} shows the result of this comparison. For all subject LLMs common between the two techniques, \name outperforms \reval in output prediction. Figure~\ref{fig:Reval} shows that while there is an overlap between the programs, these techniques correctly predicted their outputs; there are also unique cases for each technique. The number of these unique programs is higher for \name compared to \reval. We believe this is due to the well-designed prompt of \name for output prediction, with a proper in-context example and instructions to perform the \IER task.

%% file: sections/5-Related-Work.tex
\vspace{-5pt}
\section{Related Work}
\label{sec:relatedwork}

A large body of work has assessed LLMs for reasoning tasks of different modalities \cite{deshpande2021rect,wu2023reasoning,miceli2023larger,bubeck2023sparks,wang2023mathcoder,imani2023mathprompter,luo2023wizardmath,huang2023lvlms,valmeekam2022large,min2023beyond}, including natural language, visual data, math, logic, and code. \name is more closely related to the very recent studies focusing on code reasoning. 

A closely related work proposes \crux benchmark to assess the code reasoning abilities of LLMs. The dataset consists of simple programs generated by \codellama (34B) with test cases ~\cite{gu2024cruxeval}. They evaluated a series of LLMs on \crux for input and output prediction tasks. \revision{CodeSence~\cite{roy2025codesense} extends input/output prediction to reasoning problems extracted from real-world projects.}
IIP \cite{lyu2024large} proposes a novel prompting technique to enhance the accuracy of LLMs on output prediction.
\reval~\cite{chen2024reasoning} evaluates LLMs on three additional tasks: program state prediction, execution path prediction, and code coverage prediction.
Similar to \reval, \coco~\cite{beger2025coconut} challenges LLMs to generate a trace of line numbers executed by the program for a given set of inputs.
Mofia et al. \cite{la2025code} demonstrate that code execution can serve as a proxy for naturalistic tasks such as value exchange, repetitive computations, and object ranking.
\revision{CES~\cite{liu2025assessing} unifies output prediction and the reasoning on important decision points in the program within one prompt and proposes to examine LLMs' reasoning coherence and reasoning consistency.}
\revision{HoarePrompt~\cite{bouras2025hoareprompt} designs an iterative prompting technique to leverage LLMs' code reasoning capability to verify the correctness of programs with respect to their natural language specification.}
\revision{CoRe~\cite{xie2025core} further assesses LLMs' code reasoning capability on static analysis tasks, including code dependency reasoning, control dependency reasoning, and  information flow reasoning.}
Compared to prior work, \name proposes more inductive code reasoning tasks, discusses the connection between programming tasks (e.g., bug repair) and code reasoning tasks, and analyzes possible factors impacting LLMs' performance on code reasoning tasks. More importantly, \name points out the necessity of evaluating LLMs' code reasoning abilities from various aspects.

\input{Table/Table-Reval}

\name is also related to execution-aware Code LLMs, i.e., Code LLMs that are pre-trained or instruction-tuned using execution information to perform programming tasks better. NeXT~\cite{ni2024next} teaches LLMs to inspect execution traces and generate natural language rationales to reason about the run-time behavior of programs. However, NeXT is limited to its synthetic training set, which is specially designed for program repair, and can not generalize to the code reasoning tasks. \sem \cite{ding2024semcoder} instructs LLMs with operational semantics simulating the execution step by step. We evaluate \sem on \name where it even outperforms some LLMs with a larger size of parameters on some code reasoning tasks.


\vspace{-5pt}
\section{Threat To The Validity}
\label{sec:threats}


\noindent \textbf{External Validity.}
The first threat is whether our results can be generalized to other models and benchmarks. To mitigate this threat, we selected API-access (commercial) and open-access LLMs of different sizes and training strategies. We chose programs from different widely used datasets and levels of complexity to study the impact of program complexity. Our tool is publicly available to evaluate other LLMs on other datasets with different programming languages.

\vspace{3pt}
\noindent \textbf{Internal Validity.}
One potential threat to the internal validity of our results is the impact of LLMs' nondeterminism. To mitigate this threat, we used temperature $0$ to prompt all the subject LLMs. Even with temperature $0$, API-access LLMs may still show nondeterministic behavior~\cite{ouyang2023llm}, and promoting them can change the code reasoning results. 
Our results may be affected by potential bugs in implementing the \name. To address this threat, we thoroughly tested the pipeline 
and cross-checked the results for correctness. 

\vspace{3pt}
\noindent \textbf{Construct Validity.} 
\reval only studied a subset of \heval and \ceval, while we evaluated LLMs in all the programs. For common programs, the selection of tests was also inconsistent (we randomly sampled tests for output prediction, while \reval had a different number of tests). To mitigate this threat, we report the overall performance of LLMs under the \name as well as performance on the overlapped dataset.


\begin{figure}
  \begin{center}
    \includegraphics[width=0.50\textwidth]{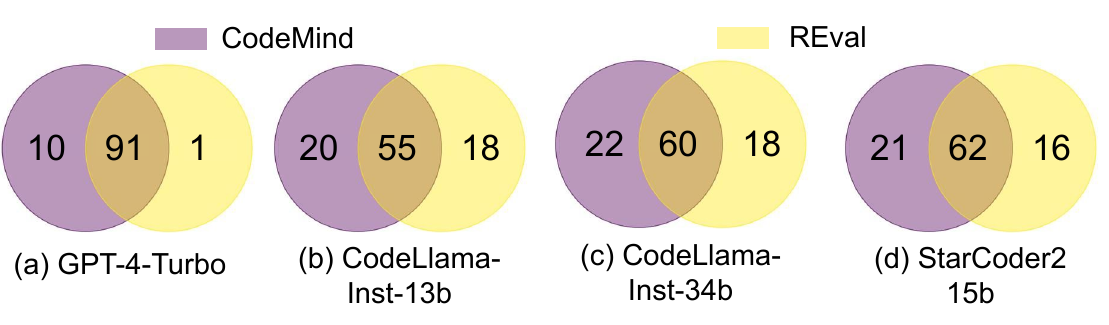}
  \end{center}
  \vspace{-5pt}
  \caption{The uniqueness and overlap between output prediction results of \name (\IER) and \reval}
  \vspace{-10pt}
  \label{fig:Reval}
\end{figure}

%% file: Table/Table-Reval.tex
\begin{table}[t]
\small
\caption{Comparison between \name and \reval's output prediction results.}
\label{tab:reval}
\centering
\vspace{-5pt}
\scalebox{0.85}{
\begin{tabular}{|l|c|c|c|c|}
\hline
\textbf{LLMs} &
  \textbf{GPT-4-Turbo} &
  \textbf{\begin{tabular}[c]{@{}c@{}}CodeLlama\\ -Inst-13b\end{tabular}} &
  \textbf{\begin{tabular}[c]{@{}c@{}}CodeLlama\\ -Inst-34b\end{tabular}} &
  \textbf{\begin{tabular}[c]{@{}c@{}}StarCoder2\\ -15b\end{tabular}} \\ \hline
\textbf{CodeMind} &
  94.39\% &
  70.09\% &
  76.64\% &
  77.57\% \\ \hline
\textbf{REval} &
  85.98\% &
  68.22\% &
  72.90\% &
  72.90\% \\ \hline
\end{tabular}
}
\vspace{-10pt}
\end{table}

%% file: sections/6-Conclusion.tex
\vspace{-5pt}

\section{Concluding Remarks }
\label{sec:conclusion}
In this paper, we discussed the necessity of code reasoning tasks as an alternative to evaluate LLMs for programming tasks. We introduced \name, a framework that supports several code reasoning tasks, and used \name in a large-scale grounded theory study to evaluate state-of-the-art LLMs for code reasoning. Our results demonstrate that LLMs, in general, know how code constructs work and achieve some levels of reasoning about program specifications. They may also follow how inputs evolve into outputs through execution. However, their ability is limited as the code becomes more complex, i.e., it has more complex control or data flow, contains non-primitive types, and invokes API calls. 

The next step for future research is to assess the code-reasoning abilities of LLMs in more realistic settings, i.e., real-world programs. This is very challenging and beyond the scope of this work, requiring (1) collecting representative programs from real-world projects and (2) proper prompt crafting and task design to enable LLMs to perform such a complex task.

\major{We provide the following key takeaways and implications for both practitioners and researchers:}

\hspace{-6pt} $\bullet$ \hspace{1pt} \major{For a comprehensive assessment, LLMs should be evaluated not only on explicit code reasoning tasks (e.g., \IER), but also on implicit code reasoning tasks (e.g., \SR and \CR).}
  
\hspace{-6pt} $\bullet$ \hspace{1pt} \major{We observe that models may succeed in program repair via reasoning shortcuts, simple pattern-matching, hallucinations, or data leakage rather than robust execution understanding, which is alarming when using them on tasks requiring reliable program analysis.}
   
\hspace{-6pt} $\bullet$ \hspace{1pt} \major{\name identified impacting factors of LLMs' performance on code reasoning, including nested code constructs, non-primitive variable types, and excessive dependencies. These findings provide guidelines for benchmark design and for curating training data and objectives that better enable in-depth code reasoning.}



\section{Acknowledgment}
This work is supported by NSF CCF 22-38045 CAR grant, IBM-Illinois Discovery Accelerator Institute, and Amazon-Illinois Center on AI for Interactive Conversational Experiences. We thank the anonymous reviewers for their comments and feedback.